\documentclass[a4paper,11pt]{article}
\usepackage{aaskaiid}
\usepackage{xcolor}
\usepackage{enumitem}
\usepackage{bm}
\usepackage{comment}
\usepackage[utf8]{inputenc}
\usepackage{orcidlink}

\usepackage{newunicodechar}
\newunicodechar{ʼ}{'}

\newcommand{\ih}[1]{{\color{blue}[Ian: #1]}}

\newcommand{\TR}{t_{\mathrm{reion}}(\Vec{r})}
\newcommand{\ZR}{z_{\mathrm{reion}}(\Vec{r})}

\definecolor{darkbrickred}{rgb}{0.55, 0.11, 0.13}

\newcommand{\hii}{{H\scalebox{0.8}{II}}}
\definecolor{cherry}{RGB}{210,4,45}

\title{Inferring Cosmology and Astrophysics from the High-redshift 21cm Signal with SKA-Low}
\ShortTitle{Parameter inference with SKA-Low}
\ShortName{Acharya et al.} 

\author[1,2]{Anshuman Acharya\orcidlink{0000-0003-3401-4884}}
\author[1,3]{Satadru Bag}
\author[4,5]{Sabrina Berger\orcidlink{0000-0002-4064-7883}}
\author[6,7]{Michele Bianco\orcidlink{0000-0002-6766-0017}}
\author[8]{Daniela Breitman\orcidlink{0000-0002-2349-3341}}
\author[9]{Lilian Crascall-Kennedy\orcidlink{0009-0009-5104-5421}}
\author[10,11]{Saswata Dasgupta}
\author[12]{Abhirup Datta\orcidlink{0000-0002-5333-1095}}
\author[9]{Marian Douspis\orcidlink{0000-0003-4203-3954}}
\author[13,14]{Ivelin Georgiev\orcidlink{0000-0002-1950-5039}}
\author[13,15]{Sambit K. Giri\orcidlink{0000-0002-2560-536X}}
\emailAdd{sambit.giri@astro.su.se}
\author[9]{Adélie Gorce\orcidlink{0000-0002-1712-737X}}
\emailAdd{adelie.gorce@cnrs.fr}
\author[16]{Caroline Heneka\orcidlink{0000-0001-8883-0583}}
\author[17]{Julien Hiegel\orcidlink{0009-0009-2235-4778}}
\author[18]{Ian Hothi\orcidlink{0000-0003-3356-5617}}
\emailAdd{ian.hothi@obspm.fr}
\author[19]{Akanksha Kapahtia\orcidlink{0000-0003-0348-0065}}
\author[20]{Nicholas Kern\orcidlink{0000-0002-8211-1892}}
\author[12]{Yashrajsinh Mahida\orcidlink{0009-0009-5104-5421}}
\author[19]{Barun Maity\orcidlink{0000-0002-4682-6970}}
\author[12]{Suman Majumdar}
\author[21]{Romain Mériot\orcidlink{0000-0003-1826-9537}}
\author[22]{Rajesh Mondal\orcidlink{0000-0001-7728-3756}}
\author[8,23]{Steven G. Murray\orcidlink{0000-0003-3059-3823}}
\author[12]{Leon Noble\orcidlink{0009-0004-3138-1130}}
\author[12]{Samit Kumar Pal\orcidlink{0000-0002-2271-4165}}
\author[24]{Aurel Schneider\orcidlink{0000-0001-7055-8104}}
\author[25]{Benoît Semelin}
\author[1,26]{Abinash Kumar Shaw\orcidlink{0000-0002-6123-4383}}
\author[27,28,29]{Hayato Shimabukuro\orcidlink{0000-0003-4850-0656}}
\author[30]{Emilie Thélie\orcidlink{0000-0001-8838-1394}}
\author[12]{Anshuman Tripathi\orcidlink{0000-0002-5091-9950}}
\author[4,31,32]{Cathryn M. Trott\orcidlink{0000-0001-6324-1766}}

\affiliation[1]{Max-Planck-Institut für Astrophysik, Garching D-85748, Germany}
\affiliation[2]{Berkeley Center for Cosmological Physics, University of California, Berkeley, CA 94720, United States}
\affiliation[3]{Technical University of Munich, TUM School of Natural Sciences, Physics Department, James-Franck-Straße 1, 85748 Garching, Germany}
\affiliation[4]{ARC Centre of Excellence for All Sky Astrophysics in 3 Dimensions (ASTRO 3D)}
\affiliation[5]{School of Physics, University of Melbourne, Parkville, VIC 3010, Australia}
\affiliation[6]{Laboratoire d'Astrophysique, École Polytechnique Fédérale de Lausanne (EPFL), Observatoire de Sauverny, Chemin Pegasi 51, CH-1290 Versoix, Switzerland}
\affiliation[7]{Institute for Particle Physics and Astrophysics, ETH Zurich, Wolfgang-Pauli-Str 27, CH-8093 Zurich, Switzerland}
\affiliation[8]{Scuola Normale Superiore (SNS), Piazza dei Cavalieri 7, I-56125 Pisa, PI, Italy}
\affiliation[9]{Université Paris-Saclay, CNRS, Institut d'Astrophysique Spatiale, 91405, Orsay, France}
\affiliation[10]{Institute of Astronomy, University of Cambridge, Cambridge, U.K.}
\affiliation[11]{Kavli Institute for Cosmology, University of Cambridge, Cambridge, U.K.}
\affiliation[12]{Department of Astronomy, Astrophysics and Space Engineering, Indian Institute of Technology Indore, Indore 453552, India}
\affiliation[13]{Department of Astronomy and Oskar Klein Centre, AlbaNova, Stockholm University, SE-10691 Stockholm, Sweden}
\affiliation[14]{ARCO (Astrophysics Research Center), Department of Natural Sciences, The Open University of Israel, 1 University Road, PO Box 808, Ra’anana 4353701, Israel}
\affiliation[15]{Van Swinderen Institute for Particle Physics and Gravity, University of Groningen, Nijenborgh 3, 9747 AG Groningen, The Netherlands}
\affiliation[16]{Institut für Theoretische Physik, Universität Heidelberg, Philosophenweg 16, 69120 Heidelberg, Germany}
\affiliation[17]{Institut Pluridisciplinaire Hubert Curien (IPHC), Université de Strasbourg, CNRS/IN2P3, 67037 Strasbourg, France}
\affiliation[18]{Laboratoire de Physique de l'ENS, ENS, Université PSL, CNRS, Sorbonne Université, Université Paris Cité, 75005, Paris, France}
\affiliation[19]{Max-Planck-Institut für Astronomie, Königstuhl 17, D-69117 Heidelberg, Germany}
\affiliation[20]{Department of Physics, University of Michigan, Ann Arbor, MI, USA, 48109}
\affiliation[21]{Department of Physics, Blackett Laboratory, Imperial College, London, SW7 2AZ, UK}
\affiliation[22]{Department of Physics, National Institute of Technology Calicut, Calicut, 673601, Kerala, India}
\affiliation[23]{Department of Physics, Stellenbosch University, Matieland 7602, South Africa}
\affiliation[24]{Department of Astrophysics, University of Zurich, Winterthurerstrasse 190, 8057 Zurich, Switzerland}
\affiliation[25]{Observatoire de Paris, PSL Research University, Sorbonne Université, CNRS, LUX, 75014, Paris, France}
\affiliation[26]{Department of Computer Science, University of Nevada, Las Vegas, NV 89154, USA}
\affiliation[27]{South-Western Institute for Astronomy Research (SWIFAR), Yunnan University, Kunming, Yunnan 650500, People's Republic of China}
\affiliation[28]{Key Laboratory of Survey Science of Yunnan Province, Yunnan University, Kunming, Yunnan 650500, People's Republic of China}
\affiliation[29]{Nagoya University, Graduate School of Science, Division of Particle and Astrophysical Science, Chikusa-Ku, Nagoya, 464-8602, Japan}
\affiliation[30]{Department of Astronomy, University of Texas at Austin, 2512 Speedway, Austin, Texas 78712, USA}
\affiliation[31]{International Centre for Radio Astronomy Research, Curtin University, 6102, Bentley WA, Australia}
\affiliation[32]{CSIRO Space \& Astronomy, Kensington, Australia}

\abstract{
The Square Kilometre Array's low frequency telescope (SKA-Low) will enable inference of astrophysical and cosmological parameters from the redshifted 21\,cm signal, probing the Cosmic Dawn and Epoch of Reionisation. While the power spectrum is the primary target for initial detection, the inherently non-Gaussian nature of the 21\,cm signal, driven by the patchy evolution of ionised regions and spin temperature fluctuations, encodes rich information accessible through higher-order statistics and morphological measurements. Extracting these constraints requires diverse inference tools, encompassing both sophisticated modelling frameworks (analytical, semi-numerical, numerical, and emulators) used to predict the 21\,cm signal, and advanced inference techniques (Bayesian, simulation-based, field-level) to connect statistics to the underlying physics. This chapter reviews these tools and explores the constraining power of different statistical probes accessible with SKA-Low, including the power spectrum, statistics beyond order two, moments of the signal distribution, and morphological measures. Combining these complementary statistics is crucial for breaking parameter degeneracies and unveiling the properties of the early Universe. We specifically assess the potential of the initial SKA-Low configuration (AA*) to measure galaxy and IGM properties, demonstrating its capability for early science results. This chapter forms part of a comprehensive set detailing the Epoch of Reionisation and Cosmic Dawn science case for the SKA-Low telescope.
}

\begin{document}
\maketitle
\newcommand{\actaa}{Acta Astron.} 
\newcommand{\araa}{ARA\&A} 
\newcommand{\aar}{A\&ARv} 
\newcommand{\aapr}{A\&ARv} 
\newcommand{\ab}{Astrobiol.} 
\newcommand{\aj}{AJ} 
\newcommand{\apj}{ApJ} 
\newcommand{\apjl}{ApJL} 
\newcommand{\apjs}{ApJSS} 
\newcommand{\ao}{Appl. Opt.} 
\newcommand{\apss}{Astro. \& Space Sci.} 
\newcommand{\aap}{A\&A} 
\newcommand{\aaps}{A\&AS.} 
\newcommand{\baas}{Bull. Am. Astron. Soc.} 
\newcommand{\caa}{Chinese A\&A} 
\newcommand{\cjaa}{Chinese J. A\&A} 
\newcommand{\cqg}{Class. Quantum Gravity} 
\newcommand{\gal}{Galaxies} 
\newcommand{\gca}{Geo. Cosmo. Acta} 
\newcommand{\icarus}{Icarus} 
\newcommand{\jcap}{JCAP} 
\newcommand{\jgr}{J. Geophys. Res.} 
\newcommand{\jgrp}{J. Geophys. Res. Planets} 
\newcommand{\jqsrt}{J. Quant. Spectrosc. Radiat. Transf.} 
\newcommand{\memsai}{Mem. SAIt} 
\newcommand{\mnras}{MNRAS} 
\newcommand{\nat}{Nature} 
\newcommand{\nastro}{Nat. Astron.} 
\newcommand{\ncomms}{Nat. Commun.} 
\newcommand{\nphys}{Nat. Phys.} 
\newcommand{\na}{New Astron.} 
\newcommand{\nar}{New Astron. Rev.} 
\newcommand{\physrep}{Phys. Rep.} 
\newcommand{\pra}{Phys. Rev. A} 
\newcommand{\prb}{Phys. Rev. B} 
\newcommand{\prc}{Phys. Rev. C} 
\newcommand{\prd}{Phys. Rev. D} 
\newcommand{\pre}{Phys. Rev. E} 
\newcommand{\prx}{Phys. Rev. X} 
\newcommand{\prl}{Phys. Rev. Let.} 
\newcommand{\psj}{Planet. Sci. J.} 
\newcommand{\planss}{Planet. Space Sci.} 
\newcommand{\pnas}{Proc. Natl Acad. Sci. USA} 
\newcommand{\procspie}{Proc. SPIE} 
\newcommand{\pasa}{PASA} 
\newcommand{\pasj}{PASJ} 
\newcommand{\pasp}{PASP} 
\newcommand{\rmxaa}{RMXAA} 
\newcommand{\sci}{Science} 
\newcommand{\sciadv}{Sci. Adv.} 
\newcommand{\solphys}{Sol. Phys.} 
\newcommand{\sovast}{Soviet Ast.} 
\newcommand{\ssr}{Space Sci. Rev.} 
\newcommand{\uni}{Universe} 

\section{Introduction}
\label{sec:intro}

The first billion years of cosmic history, spanning the Cosmic Dawn (CD; the era of the first stars, $z\sim 30$–$15$) and the subsequent Epoch of Reionisation (EoR; when the intergalactic medium became fully ionised, $z\sim 15$–$6$), are central to astrophysics and cosmology. Key questions include the properties of the first stars and galaxies, how their radiation set the thermal and ionisation state of the intergalactic medium (IGM), and how this evolution traced the growth of large-scale structure. The redshifted 21\,cm signal from neutral hydrogen is a powerful probe of this era: because its observed frequency maps directly to redshift, it enables three-dimensional (tomographic) mapping of the IGM \citep[for a review, see e.g.][]{greig_2019}. Radio telescopes such as the low-frequency component of the Square Kilometre Array (SKA-Low) measure the 21\,cm differential brightness temperature relative to a radio background,
\begin{align}
    \delta T_{\mathrm{b}} (\bm{r}, z) = 27 \mathrm{mK}\left(\frac{1+z}{10}\frac{0.27}{\Omega_{\mathrm{m}}}\right)^{1/2} \left( \frac{\Omega_{\mathrm{b}}}{0.044} \frac{h}{0.7} \right) x_{\rm HI}(\bm{r})\left[1+\delta_{B}(\bm{r})\right] \left( \frac{T_{\mathrm{s}}(\bm{r},z) -T_{\mathrm{R}}(z) }{T_{\mathrm{s}}(\bm{r},z)} \right)
    \label{eq:dTb}
\end{align}
where $\delta_B$ is the baryonic overdensity, $x_{\rm HI}$ the neutral hydrogen fraction, and $T_{\mathrm{s}}$ and $T_{\mathrm{R}}$ the spin and radio-background temperatures, respectively. In the standard model this background is the cosmic microwave background (CMB), although some studies suggest deviations \citep[e.g.,][]{Fixsen_2011,Dowell_2018}. In the high spin-temperature regime ($T_{\mathrm{s}} \gg T_{\mathrm{CMB}}$) the signal is insensitive to spin-temperature fluctuations, simplifying its interpretation. The first statistical detection of the 21\,cm signal is expected from the spherically averaged power spectrum, which can constrain the properties of early ionising sources \citep[e.g.,][]{FialkovBarkana_2014,QinMesinger_2021,munoz2023effective}, the timing and duration of reionisation \citep[e.g.,][]{Maity_Choudhury_2023, GharaShaw_2024}, and fundamental cosmological parameters \citep[e.g.,][]{Heneka:2018ins, Liu2020, GiriSchneider_2022, SchneiderSchaeffer_2023}.

Although the redshifted 21\,cm signal has not yet been detected, pathfinder experiments set tight upper limits on its power spectrum. LOFAR, from $\sim$200\,h of data, finds $\Delta^2_{21,{\rm UL}}$ below $(68.7\,{\rm mK})^2$, $(54.3\,{\rm mK})^2$, and $(65.5\,{\rm mK})^2$ at $z\approx10.1$, $9.1$, and $8.3$, respectively, giving the strongest constraints near $k \approx 0.076\,h\,{\rm cMpc}^{-1}$ \citep{MertensMevius_2025}. At slightly lower redshifts, MWA (268\,h) yields $\Delta^2_{21,{\rm UL}}$ below $(23.0\,{\rm mK})^2$, $(30.5\,{\rm mK})^2$, and $(34.9\,{\rm mK})^2$ at $z=6.5$, $6.8$, and $7.0$, at $k \approx 0.18\,h\,{\rm cMpc}^{-1}$ \citep{TrottNunhokee_2025}, while HERA gives $\Delta^2_{21,{\rm UL}}$ below $(21.4\,{\rm mK})^2$ at $z=7.9$ ($k \approx 0.34\,h\,{\rm cMpc}^{-1}$) and $(59.1\,{\rm mK})^2$ at $z=10.4$ ($k \approx 0.36\,h\,{\rm cMpc}^{-1}$) \citep{Hera2023}. Together these limits constrain the thermal and ionisation history of the IGM: a completely cold IGM is disfavoured, since LOFAR limits at $z = 9.1$ imply $T_{\mathrm{s}} > T_{\mathrm{CMB}}$, and newer HERA measurements at other redshifts and scales rule out cold reionisation scenarios \citep{GharaGiri_2020,Hera2023}. Combined with MWA upper limits at $z = 6.5$, they favour models in which reionisation was early at $z \sim 9$ and nearly complete by $z \sim 6.5$, requiring substantial heating throughout \citep{GharaGiri_2020,Hera2023}. Phase I HERA limits at $z = 8$ further indicate that the IGM had been heated above the adiabatic threshold, providing the first quantitative lower bounds on X-ray heating from early galaxies \citep{Hera2021}. They also disfavour excessive heating, large ionised regions, and strong radio backgrounds: LOFAR excludes models with large ionised regions or high heated-gas volume fractions at $z \approx 9$ and, across $z = 8.3$–$10.1$, extreme heating or reionisation scenarios driven by rare, luminous sources that produce large ionised bubbles \citep{GharaGiri_2020}; LOFAR limits also bound any excess high-redshift radio background, allowing a modest excess but ruling out the strong backgrounds implied by ARCADE2 or LWA1 \citep{2020:MondalFialkov}, while HERA constraints on radio–X-ray luminosity combinations can exclude radio-bright, X-ray-faint galaxies \citep{Hera2021}.

As reionisation is inherently non-Gaussian, the spatial distribution of the 21\,cm signal encodes higher-order and morphological information \citep{Abinash_infer2020}, which can be extracted with complementary analyses such as the bispectrum, field-level statistics, morphological tracers, and Fourier phase correlations. The first upper limits on a beyond-the-power-spectrum statistic were reported by \citet{2019PASA...36...23T} and \citet{2025_Gill}, who applied the bispectrum to MWA observations at $z=6.2$–$7.5$ and $z = 8.2$, respectively; the latter used a multi-frequency angular bispectrum estimator to separate the foreground wedge from the EoR window. Cross-correlating the 21\,cm signal with other observables—including the patchy kinetic Sunyaev–Zel’dovich effect \citep{GeorgievGorce_2024}, galaxy surveys \citep{GagnonHartman_2025}, the Ly$\alpha$ forest \citep{Qin2024}, and line-intensity mapping surveys \citep{Heneka:2017, FronenbergLiu_2024}—will further break degeneracies and improve astrophysical and cosmological inferences; see \citet{Chakraborty01.2026.SKA} for a detailed discussion.

SKA-Low will have much higher sensitivity than current facilities, enabling precision measurements of Gaussian and non-Gaussian aspects of the 21\,cm signal. Its dense core, long baselines, and flexible observing modes will let it probe a wide range of spatial scales and constrain the properties of the first galaxies, the thermal history of the IGM, and possible extensions to the standard cosmological model. This chapter reviews the current status of the field, the astrophysical and cosmological inference capabilities of SKA-Low, and synergies with other probes. We first describe modelling frameworks (Sec.~\ref{sec:modelling}), then the expected improvement in parameter constraints with SKA-Low using power spectrum and higher-order statistics (Secs.~\ref{sec:PS_inf} and \ref{sec:beyond_PS}), and strategies for extracting maximum information from 21\,cm data (Sec.~\ref{sec:max_info}). We then discuss caveats (Sec.~\ref{sec:discussion}), conclusions (Sec.~\ref{sec:conclusion}), author contributions (Sec.~\ref{sec:author_contribution}), and the Fisher inference framework (Appendix~\ref{sec:fisher_framework}).

\section{Modelling frameworks}
\label{sec:modelling}

A parameter inference framework provides physical constraints by confronting observations with models.
Several such models exist, which we review in this section, whether they are analytical (Sec.~\ref{sec:analytical}), semi-numerical (Sec.~\ref{sec:sem_num}), numerical (Sec.~\ref{sec:num}), or machine learning-based (Sec.~\ref{sec:emu}). All these methods aim to calculate the propagation of ionising photons to quantify the evolution of the hydrogen neutral fraction, $x_{\rm HI}(\bm{r})$, and the gas temperature of the IGM for the spin temperature calculation, $T_\mathrm{s}$, that are the key ingredients quantifying the 21\,cm signal during CD and EoR (Eq.~\ref{eq:dTb}). 

\subsection{Analytical}\label{sec:analytical}

The first attempts to model the 21\,cm power spectrum during CD and EoR relied on analytical methods. These methods use a combination of perturbation theory and excursion-set modelling for the ionising sources \citep{Barkana:2004vb,furlanetto2004growth, Pritchard:2006sq}. The state of the intergalactic gas (e.g., its temperature or ionisation fraction) at a specific location was calculated by summing the radiative influence of surrounding sources, accounting for their distances. This approach provided a first qualitative understanding of the physical processes during the EoR. However, they have since then been shown to deviate from numerical calculations by a factor of a few \citep{Santos_2008, SchneiderGiri_2021}.

More recently, bias expansion \citep[e.g.,][]{GiriSchneider_2022, SchneiderSchaeffer_2023} has provided an empirical connection between the 21\,cm and dark-matter power spectra \citep[e.g.,][]{GeorgievGorce_2024, GharaShaw_2024} by calibrating on simulations. In parallel, \citet{mirocha2022galaxy} developed a phenomenological model of the EoR 21\,cm power spectrum that works directly in terms of IGM properties, without relying on complex galaxy and source models, whilst assuming a bubble size distribution (see Sec.~\ref{sec:bubble_stats}). In another approach, \citet{munoz2023effective} built a model for the pre-reionisation 21\,cm signal based on an exponential parameterisation of star formation rate density. \citet{RasteSethi_2018,RasteSethi_2019} studied the two point correlations of fluctuations in the CD 21\,cm signal using a simple two-source model, including Lyman-$\alpha$ coupling. Indeed, Lyman-$\alpha$ photons play a very important role in shaping the signal during the CD. \cite{HolzbauerFurlanetto_2012} provided an analytical model for the fluctuations in Lyman-$\alpha$ and Lyman-Werner backgrounds. 

While the above works provide a fast way to derive the 21\,cm power spectrum--a significant advantage for computationally intensive parameter inference--they are limited in applicability and precision. For example, \citet{SchaefferGiri_2024} identified limitations in the perturbative expansion of the 21\,cm signal used in several analytical frameworks. See e.g., \citet{Santos_2008} for more discussion. 

\subsection{Semi-Numerical}\label{sec:sem_num}

Semi-numerical methods bridge the gap between the physical detail of numerical simulations and the computational efficiency needed for parameter inference. While numerical radiative transfer (RT) simulations capture complex small-scale physics, they are too expensive to run on SKA-like cosmological volumes or to generate the $>10^4$ realisations required for parameter estimation. Semi-numerical approaches provide a practical compromise: they model the non-linear matter distribution representing large-scale structure in large volumes while approximating the small-scale astrophysical processes during the EoR and CD through calibrated prescriptions.

The most common semi-numerical simulations are based on the excursion set formalism \citep{Bond91}, adapted to identify a region's ionisation state by comparing the cumulative ionising photon count to the number of neutral hydrogen atoms within the same volume \citep{furlanetto2004growth}. This approach progressively smooths the density field on smaller scales until the ionising photon criterion is met, efficiently capturing the inside-out nature of reionisation where over-dense regions ionise first. These models require the underlying baryon distribution and the location and mass of collapsed objects, which are obtained by modelling cosmological structure formation either through Lagrangian perturbation theory or full $N$-body simulations
followed by halo identification. The ionised field is then converted to the 21\,cm brightness temperature field, with peculiar velocities incorporated either from $N$-body simulations or via perturbative approaches.

Numerous codes implement excursion set methods, including {\sc Sem-Num} \citep{2009Choudhury, 2012Majumdar, 2013Majumdar}, {\sc SimFast21} \citep{Santos_2010}, {\sc 21cmFAST} \citep{21cmFAST_2011,Murray2020, Davies2025}, {\sc ReionYuga} \citep{Mondal_2017}, {\sc SCRIPT} \citep{choudhury_paranjape_2018,Maity_choudhury_2022}, and {\sc CIFOG} \citep{Hutter_2018}. When calibrated against full RT simulations, these methods reproduce large-scale morphology and statistics such as the power spectrum to within $\lesssim 10\%$ \citep{Zahn2011Comparison, Majumdar_2014, Ghara2018RT1D}. Modern implementations have incorporated increasingly realistic physics, including inhomogeneous X-ray heating, Lyman-$\alpha$ coupling during the cosmic dawn, inhomogeneous recombinations, and radiative feedback on ionising sources \citep{21cmFAST_2013,Maity_choudhury_2022}. Several pipelines also include semi-analytic galaxy models during reionisation, enabling joint constraints from galaxy and 21\,cm observables, such as {\sc ASTRAEUS} \citep{Hutter_2021,Hutter_2024}, {\sc MERAXES} \citep{Mutch_2016,Balu_2023} and {\sc POLAR} \citep{Ma_2023,Ma_2025,Acharya_2025}. 

Beyond excursion set approaches, alternative semi-numerical methods exist. Hybrid schemes combine density and photon fields with 1D radiative transfer calculations around each source, capturing radial ionisation and thermal profiles at modest computational cost. Examples include {\sc BEARS} \citep{Thomas_2009}, {\sc GRIZZLY} \citep{Ghara_2015,Ghara2018RT1D}, and {\sc BEoRN} \citep{Schaeffer_2023}. More recently, effective field theory (EFT)-based approaches have been developed to model the large-scale 21\,cm field and power spectrum using bias expansions that systematically incorporate small-scale astrophysics \citep{McquinnDAloisio_2018}. These frameworks have been implemented on simulation grids \citep[e.g.,][]{QinSchutz_2022, BaradaranHadzhiyska_2024}, offering a complementary approach to excursion set methods. Additionally, some codes such as {\sc AMBER} \citep{Trac_2022} and {\sc ARTIST} \citep{Molaro_2019} employ alternative methodologies for modelling the reionisation process. Thus, the computational efficiency of semi-numerical methods makes them indispensable for the parameter estimation from SKA observations, either through direct use in inference frameworks or by generating large datasets for training emulators (Sec.~\ref{sec:emu}).

\subsection{Numerical}\label{sec:num}

Numerical simulations provide the most physically detailed approach to studying the CD-EoR. These frameworks differ fundamentally in how they model structure formation: some employ full hydrodynamical simulations that self-consistently evolve the large-scale structure and ionising source formation, while others use computationally cheaper dark matter-only simulations combined with recipes to model ionising sources within dark matter haloes. Regardless of the approach, these simulations solve the 3D RT equation to track how photons propagate from their sources and impact the IGM, coupled with ordinary differential equations (ODEs) that describe the resulting ionisation and thermal state of the gas through primordial chemistry and heating/cooling processes.

\begin{figure}[!ht]
    \centering
    \includegraphics[width=\textwidth]{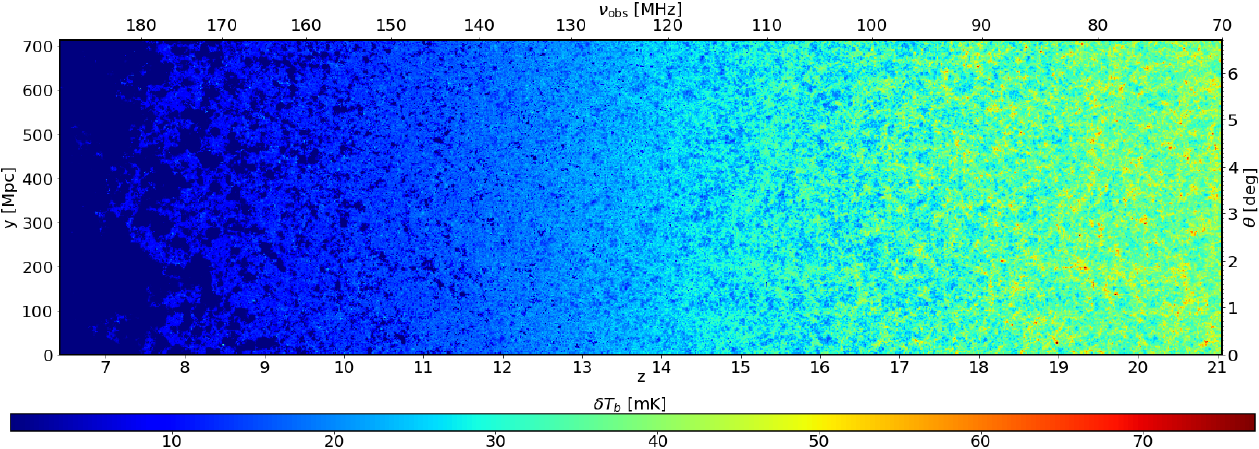}
    \caption{A slice through the 21\,cm tomographic lightcone (LC) simulated with the {\sc C$^2$Ray} (numerical code) showing the signal evolving between $z=21$ and $6$ ($\nu_\mathrm{obs}=70-190\,\mathrm{MHz}$). The simulation has a comoving angular length of $\sim 700\,\mathrm{Mpc}$, corresponding to a field of view of $\sim 4.5$ degrees at $z=7$. 
    }
    \label{fig:lc_compare}
\end{figure}

Methods for solving the RT equation during reionisation fall into two broad classes: \textit{Moment-based} and \textit{Ray-tracing} schemes. Moment-based methods approximate the radiation field by evolving its angular moments (energy and flux) using the same computational mesh as the hydrodynamics, naturally coupling radiation with gravity and fluid dynamics. This maintains consistency between gas dynamics and radiative processes, but becomes computationally expensive even for volumes as small as $\sim100$ cMpc \citep{Garaldi2022, Ocvirk2016codaI, Ocvirk2020codaII}.
Example frameworks include {\sc ATON} \citep{Aubert2010atonrt,Asthana2024atonhe}, {\sc RAMSES-RT} \citep{Rosdahl2015} and {\sc AREPO-RT} \citep{Kannan2019areport}.
Ray-tracing algorithms model rays of photon as they propagate through the IGM.
They provide high directional accuracy, photon conservation, and more physical treatment of shadowing and anisotropic radiation, making them well-suited for resolving small-scale ionisation structures \citep[e.g.,][]{Iliev2006ComparisonI}.
Example codes include {\sc CRASH} \citep{Ciardi2001MCRT}, 
{\sc pyC$^2$Ray} \citep{Mellema2006c2ray,Hirling2024pyc2ray}, {\sc Zeus-MP} \citep{Whalen2006Zeus}, {\sc SPH-RAY} \citep{Altay2008sphray}, {\sc LICORICE} \citep{Baek09,Meriot25} and {\sc FlexRT} \citep{Cain2024FlexRT}.

Numerical simulations remain the most physically detailed tool for modelling the cosmic dawn and reionisation, capturing the complex interplay between structure formation, radiative transfer, and gas thermodynamics. Several publicly available simulation databases have been developed to support EoR studies, such as the \textsc{Thesan}\footnote{\url{https://www.thesan-project.com/}} \citep{Smith2022, Garaldi2022, Borrow2023, Garaldi2024thesandata}, \textsc{LoReLi}\footnote{\url{https://21ssd.obspm.fr/}} \citep{Meriot24, Meriot25}, and \textsc{StoReS}\footnote{\url{https://github.com/sambit-giri/StoReS}} \citep{Dixon2015, GiriBianco2024}. Fig. \ref{fig:lc_compare} shows an example 21\,cm lightcone (LC) simulated using {\sc C$^2$Ray}, described in \citet{Bianco2021inhomo}, illustrating the signal evolution from $z=21$ to 6. However, the high computational cost of numerical simulations limits their direct use in Bayesian inference. Instead, they play a crucial role in calibrating the semi-numerical frameworks (Sec. \ref{sec:sem_num}) and training the emulators discussed in subsequent sections (Sec.\ref{sec:emu}), which enable efficient parameter estimation from SKA observations.

\subsection{Emulators}
\label{sec:emu}

Modern cosmological parameter inference approaches using Markov Chain Monte Carlo (MCMC) require $O(10^{6})$ model evaluations \citep[e.g.,][]{GreigMesinger_2015}. This is computationally prohibitive even for the fastest semi-numerical codes (Sec.~\ref{sec:sem_num}), which take $O(1)$ CPU-hr per simulation. Emulators, or surrogate models, solve this by learning a fast mapping from parameters to observables, reducing a single forward model evaluation to O(CPU sec) \citep[e.g.,][]{Heitmann2006, Heitmann2009}. The computational cost of inference is thus significantly \textit{amortised}: the primary effort is the initial, one-time cost of producing a training database, followed by training and testing the emulator (often a Gaussian Process or neural network; see \citealt{Acharya02.2026.SKA} for more details about emulator architectures). This allows a single trained emulator to support many inferences on different datasets \citep[e.g.,][]{GharaGiri_2020,BreitmanMesinger_2024}.

Two key challenges in building an emulator are generating an efficient training set and validating its fidelity. The number of samples needed to cover a parameter space scales exponentially with its dimensionality. This ``curse of dimensionality'' can be mitigated with smart sampling methods, such as Latin Hypercubes \citep[e.g.,][]{schmit2018emulation,Meriot24}, or by focusing the training set on regions of interest, for instance, by re-using posterior samples from previous analyses (e.g., \citealt{GorceDouspis_2022,BreitmanMesinger_2024}).
After training, the emulator must be validated. This is typically done with a hold-out test set—a subset of simulations withheld from training. A common standard is to require the emulator's prediction error to be significantly smaller ($\sim 10\%$) than the observational error \citep{Bevins2025}. If the emulator error is non-negligible, its covariance matrix should be included in the likelihood to prevent biased posterior distributions.

Emulators of the 21 cm signal can be split into two broad categories. The first and most common type predicts low-dimensional summary statistics, such as the 21 cm PS \citep{kern2017emulating, schmit2018emulation, GharaGiri_2020, TiwariShaw_2022, BreitmanMesinger_2024, Meriot24, MaityParanjape_2023, Choudhury_Paranjape_Maity_2024, 2025Tripathi,maity_2025, Choudhury_2025, Mahida_2025}, global signal \citep{cohen2020emulating, globalemu, VAE, BreitmanMesinger_2024,2024LSTM}, and bispectrum \citep{TiwariShaw_2022, Mahida_2025}. These are computationally cheaper to train and can achieve sub-percent accuracy, enabling synergistic inferences with multiple probes \citep{BreitmanMesinger_2024}. The second category consists of field-level emulators that produce full 21 cm maps \citep[e.g.,][]{Zhao_2023, Heneka_2025, Mishra25} or neutral fraction \citep[e.g.,][]{Chardin2019, Korber2023PINION, Posture_2025}. While vital for field-level inference (Sec. \ref{sec:fieldlevel}), these are far more computationally expensive to train and may struggle to accurately learn features across all scales.

\section{Constraining power of the power spectrum}
\label{sec:PS_inf}

With the modelling frameworks established above, we now turn to parameter inference. This section focuses on the 21 cm power spectrum (PS), the primary summary statistic for initial detection. We first discuss different formulations of the PS and their forecast constraining power in Sec. \ref{sec:PS}. In Sec. \ref{sec:inf_comp}, we review various Bayesian inference techniques used to derive parameter constraints.

\subsection{Power spectrum}

\label{sec:PS}

The PS is the primary statistic that all experiments are currently attempting to detect to characterise the spatial fluctuations of the 21 cm signal. 
The PS is important in cosmology because, for a stationary Gaussian random field—a good approximation to many cosmological fields on appropriate scales and redshifts—it provides a statistically complete description.

This section reviews various forms of the PS, including the spherically-averaged (1D) and cylindrically-averaged (2D) PS in Sec.~\ref{subsec:ps}, and the Multi-frequency Angular Power Spectrum (MAPS) in Sec.~\ref{subsec:maps}. We discuss the results of a forecast study with SKA-Low observations -- considering the AA* and AA4 layouts, to determine the astrophysical constraints we expect to reach with the SKA in Sec.~\ref{sec:PS_fisher}.

\subsubsection{Spherically- and cylindrically-averaged power spectrum}\label{subsec:ps}

The 21\,cm PS is defined as \citep[e.g., ][]{2025PASP..137g3001M}:
\begin{align}
\langle \tilde{T}_\mathrm{b}(\bm{k})\tilde{T}_{\mathrm{b}}^\ast(\bm{k}^\prime)\rangle = (2\pi)^3 P(\bm{k})\delta^D(\bm{k}-\bm{k}^\prime),
\end{align}
where $P(\bm{k})$ is the 3D PS at wave-vector $\bm{k}$, $\tilde{T}_\mathrm{b}(\bm{k})$ is the 3D Fourier transformed 21\,cm differential brightness temperature field $\delta T_{\rm b}(\bm{r})$ (Eq.~\ref{eq:dTb}), $\delta^D$ is the Dirac delta function, and $\langle\rangle$ denotes an ensemble average.
Thus, the PS is the expectation value of the squared Fourier-transformed field. Under isotropy and homogeneity—valid for a \emph{coeval} volume\footnote{A comoving field at fixed $z$, in contrast to a lightcone, which includes redshift/time evolution along one dimension.}—$P$ depends only on the magnitude $k=|\bm{k}|$, so the single function $P(k)$ captures much of the field’s statistical information.

Telescopes, however, observe a \textit{lightcone} (LC) in which the Universe evolves along the line of sight. It is therefore useful to distinguish line-of-sight modes $k_\parallel$ from transverse modes $\bm{k_\perp} \equiv (k_{\perp, x}, k_{\perp,y})$. Within a 3D LC volume represented on a regular mesh, an \textit{estimate} of the 3D PS is obtained by taking a 3D Fourier transform and squaring. Averaging modes in cylindrical bins of constant $k_\perp \equiv  |\bm{k_\perp}|$ yields the 2D (cylindrically averaged) PS, while averaging in spherical shells of constant $k$ yields the 1D (spherically averaged) PS. Both are \textit{estimates}, not ensemble averages, and the evolving line-of-sight axis introduces the \emph{LC effect} (see Sec.~\ref{subsec:maps}). In Fig.~\ref{fig:PS_fiducial}, we show the 1D (left) and 2D (right) PS of a simulated LC (see Appendix~\ref{sec:fisher_simulations} for simulation details).

Real observations are less ideal, and PS estimators are more generally described by the \textit{quadratic estimator formalism} \citep{LiuParsons_2014a,LiuParsons_2014b,DillonNeben_2015,LiuShaw_2020}, which expresses the estimate as a quadratic form in the visibility data, naturally incorporating choices such as binning, Fourier transforms, filters and weights. Within this framework, the observed PS estimate is related to the `true' PS via: 
\citep[e.g.,][]{GorceGanjam_2023,FronenbergLiu_2024}:
\begin{equation}\label{eq:ps_w_wf}
\hat{P}(k) =  \int \mathrm{d} k'  \,P_\mathrm{true}(k')\,W(k, k') \ ,
\end{equation}
where the \textit{window function} $W$ encodes the instrument, observational setup, and analysis pipeline, and must be included in forward models when inferring physics.

As the cosmological 21\,cm signal is nearly isotropic, 1D and 2D PS estimates contain comparable cosmological information and lead to similar parameter constraints \citep{GreigPrelogovic_2024}. In detail, however, the 2D PS retains additional line-of-sight information from observational effects such as redshift-space distortions and the Alcock--Paczy{\'n}ski effect \citep[see Sec.~\ref{sec:PS_fisher} and][]{BarkanaLoeb_2005,Barkana_2006,LiuTegmark_2011,PoberLiu_2014,JensenDatta_2013}.

The main advantage of working in cylindrical $(k_\perp,k_\parallel)$ space is the ability to model foregrounds and systematics more naturally in the forward model \citep{Kern_2025}. Because the line of sight is traced by frequency (via its relation to the 21\,cm redshift), non-cosmological components such as foregrounds behave very differently along this axis than across the sky: their extremely smooth frequency evolution confines them to the largest-scale modes after a Fourier transform. The right panel of Fig.~\ref{fig:PS_fiducial} shows this foreground contamination in the 2D PS. The white lines mark the `foreground wedge' \citep{Datta2010,LiuParsons_2014a,Murray2018}, the region dominated by intrinsically smooth-spectrum foreground emission. Its slope can be predicted geometrically \citep{Parsons2012} from assumptions about the largest zenith angle at which foregrounds contribute (after primary-beam attenuation). Two such assumptions are illustrated: the horizon limit (dashed), which assumes foregrounds extend to the horizon, and the `primary-beam wedge' (solid), which assumes a sharp decline outside the main lobe of the beam \citep[e.g.,][]{PoberLiu_2014}. The region outside the wedge is the \emph{21\,cm window}, where the cosmological signal is typically sought. The corresponding contaminated region in the 1D PS is hatched in the left panel and almost overlaps the noise-dominated region for 100~hours of AA* observations, illustrating the need for foreground removal to recover the 21\,cm signal inside the horizon wedge. Further discussion of foreground contamination in the high-redshift 21\,cm signal can be found in \citet{Burba01.2026.SKA}.
\begin{figure}[htbp]
    \centering
    \includegraphics[width=0.46\textwidth]{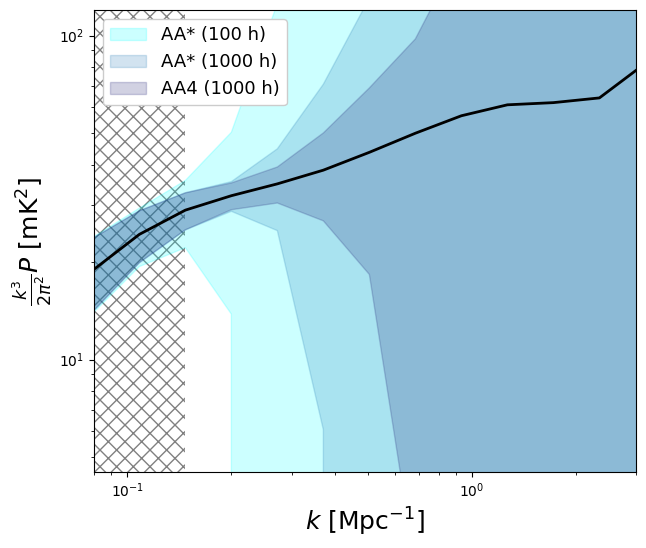}
    \includegraphics[width=0.46\textwidth]{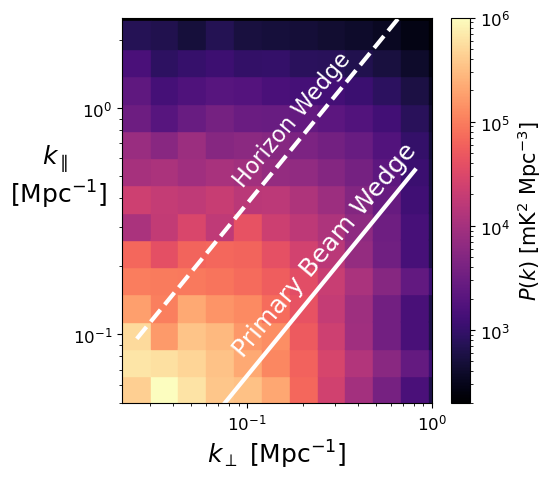}
    \caption{
        Power spectra (PS) of the fiducial model used in our Fisher forecast (Appendix~\ref{sec:fisher_framework}).
        \textit{Left:} The spherical PS, $P(k)$. The shaded areas correspond to the forecasted $1\sigma$ errors, which include sample variance and thermal noise, for three SKA-Low scenarios: AA* (100 h), AA* (1000 h), and AA4 (1000 h). The hatched region corresponds to foregorunds-dominated wavemodes (wedge).
        \textit{Right:} The 2D cylindrical PS, $P(k_\perp, k_\parallel)$. The lines illustrate the horizon-limited (dashed) and an optimistic primary beam-limited (solid) `foreground wedge'. The region outside this wedge is the `EoR window', where the signal is typically sought.
    }
    \label{fig:PS_fiducial}
\end{figure}

\subsubsection{Multi-frequency angular power spectrum} \label{subsec:maps}

A key challenge arises because the 21\,cm signal evolves along the line-of-sight (LoS), due to cosmic evolution -- the LC effect, making the signal statistically non-ergodic along the LoS, and breaking the statistical isotropy assumption underlying the spherical power spectrum definition.
Simulations have shown that directly estimating a 1D PS from a LC segment can yield estimates that differ from $P(k, z)$ defined at a \textit{coeval} volume at the central $z$ of the segment by more than 100\% in some cases, depending on the epoch and the rate of evolution (e.g., \citealt{Kanan_lightcone2012, Kanan_lightcone2014, LaPlante_lightcone2014, Rajesh_lightcone2018, Abinash_lightcone2023}). 

Typically, most precursors manage the LC effect by computing the power spectrum over sufficiently limited bandwidths. 
Alternative statistics have been developed to take the LC effect into account, e.g., using a spherical harmonic basis \citep{Kanan_2007} or a wavelet basis \citep{2016MNRAS.461..126T}.
\citet{Rajesh_lightcone2018} proposed the Multi-frequency Angular Power Spectrum (MAPS), a two-point non-stationary statistic which correlates the 2D Fourier-transformed signal on the sky plane at two different frequencies $\nu_1$ and $\nu_2$. Under the flat-sky approximation, the MAPS can be written 
\begin{equation}
    \mathcal{C}_{\ell}(\nu_1, \nu_2) = \Omega^{-1} \,\langle \tilde{T}_{{\rm b} 2}(\bm{U}, \nu_1)\, \tilde{T}_{{\rm b} 2} (-\bm{U}, \nu_2) \rangle~,
    \label{eq:cl}
\end{equation}
where $\bm{U}$ is the baseline vector, $\ell = 2\pi |\bm{U}|$ is the angular multipole, $\tilde{T}_{{\rm b} 2}(\bm{U}, \nu)$ is the 2D Fourier transform of the brightness temperature field $\delta T_{\rm b}(\bm{r}, z)$ on the sky plane, and $\Omega$ is the solid angle subtended by the observed (or simulated) sky-patch. Interestingly, unlike the standard PS, the MAPS does not require a Fourier transform along the frequency axis. Because the MAPS performs less averaging than the spherical power spectrum, the former retains more information but leads to comparatively lower signal-to-noise ratios \citep[SNR, see ][for a proposed solution]{Pramanick2025}. 
SKA-Low is projected to measure the diagonal elements of the 21\,cm EoR MAPS (auto-frequency) at intermediate scales ($\ell \sim 1000$) with $1000$ hours of data at $\gtrsim  3\sigma$ in foreground-free scenarios \citep{Rajesh_lightcone2020} -- the scenario we consider in our forecasts (Sec.~\ref{sec:PS_fisher}). During CD, the 21\,cm MAPS is orders of magnitude larger than during the EoR, such that $100$~hours of observation are sufficient to reach the same detection level on $300 \lesssim \ell \lesssim 2000$ \citep{Abinash_lightcone2023}. Accounting for foreground avoidance worsens the SNR but detection is still achievable at specific redshifts and scales \citep{Rajesh_lightcone2020}.

\subsubsection{A simple Fisher forecast} \label{sec:PS_fisher}

\begin{figure}
    \centering
    \includegraphics[width=0.48\textwidth]{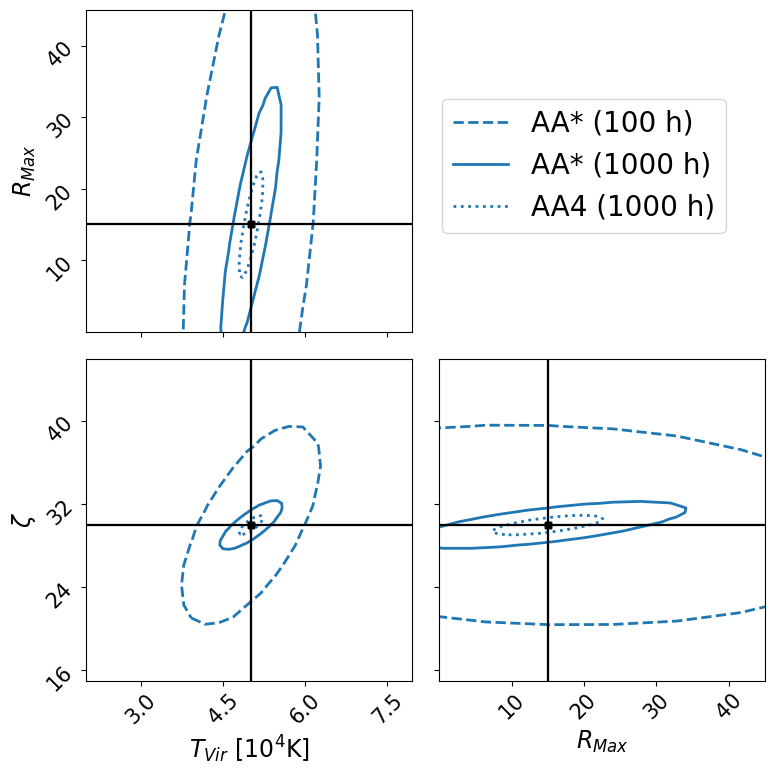}
    \includegraphics[width=0.48\textwidth]{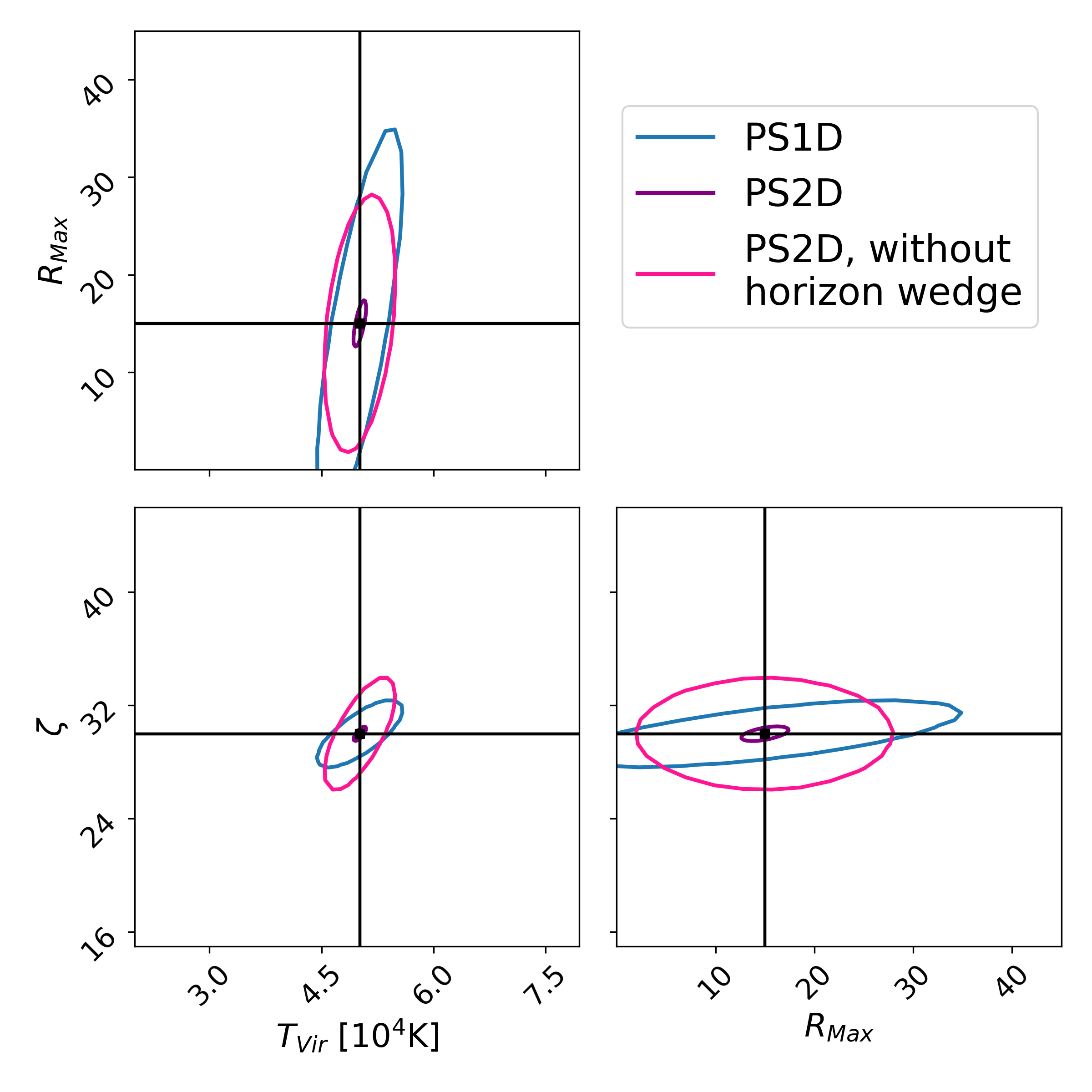}
    \vspace{-1em}
    \caption{
        Posterior distributions from a Fisher forecast study with mock SKA-Low data.
        \textit{Left:} 
        Constraints from the spherical power spectrum (PS) for three scenarios: AA* 100h (dashed blue), AA* 1000h (solid blue), and AA4 1000h (dotted blue), show improving constraints when either the observation time increases or the SKA-Low antenna configuration becomes denser.
        \textit{Right:} A comparison of constraints from the spherical PS (solid blue) and cylindrical PS, with (solid purple) and without (solid pink) including the horizon wedge, all for the AA* 1000 h survey.
        }
    \label{fig:PS_Corner_plot}
\end{figure}

To evaluate the scientific capability of SKA-Low, we perform a Fisher forecast for two key construction stages: AA* (307 antennas) and AA4 (512 antennas). This forecast is performed at $z\approx 8$ 
for the three forms of the PS defined above, and for three astrophysical parameters:
(i) the minimum virial temperature $T_\mathrm{vir}=5\times10^4$\,K, (ii) the maximum allowed distance for ionising photon transfer $R_\mathrm{max}=15$ Mpc, and (iii) ionising photon production efficiency $\zeta=30$. The data covariance matrix was estimated from an ensemble of 400~simulations at the fiducial cosmology to account for sample variance and instrumental noise (see Appendix~\ref{sec:fisher_obs_setup} for details, including our 21\,cm signal and instrumental models). Its diagonal terms (i.e., the variance in each $k$-bin) make the error bars in the left-hand panel of Fig.~\ref{fig:PS_fiducial}. 
Note that a Fisher analysis yields the best constraints possible \citep[e.g.,][]{Heavens_2009}, obtainable by an efficient estimator (which may not exist). Here, we use it to estimate the relative improvement between different observational setups, and to compare the information content of different statistics.

The left panel of Fig.~\ref{fig:PS_Corner_plot} presents the 2-$\sigma$ contours obtained with the spherical PS. The constraints improve significantly with longer integration times (from 100 hours to 1000 hours with AA*) and with the more powerful AA4 array, highlighting a clear path from early science to precise parameter estimation with the full SKA-Low observatory.

On the right panel of Fig.~\ref{fig:PS_Corner_plot}, we show that the cylindrical PS can provide better constraints on the astrophysics than the spherical PS, as long as the wedge-modes are recovered behind the foregrounds \citep[see also][]{Breitman2025}. \textbf{Note that Fisher contours are a best-case scenario, and the improvement in constraining power between the spherical and the cylindrical power spectrum} can vary when performing Bayesian parameter inference \citep{GreigPrelogovic_2024}. Additionally, the cylindrical basis is useful for visualising the impact of systematics, such as foregrounds, and applying avoidance strategies -- excising contaminated modes. Applying such cuts (whether subsequently binned into 1D or not) degrades the final constraints by a factor of 2-3 due to information loss \citep[see also][]{GreigPrelogovic_2024}. Conversely, MAPS provides a way to potentially increase the information extracted. While the MAPS auto-correlation components ($\mathcal{C}_{\ell}(\nu, \nu)$) yield constraints comparable to the 1D spherical PS, constraints could improve by a few factors if the cross-frequency (off-diagonal) components are included.

\subsection{Comparing inference approaches with the power spectrum}
\label{sec:inf_comp}

We employed a Fisher analysis in the previous sub-section for its ease of use. However, Fisher analysis provides best-case constraints and more elaborate approaches must be used to provide accurate forecasts of the science capabilities of the SKA. In this sub-section, we review more elaborate parameter inference frameworks that will be applied to the upcoming SKA-Low data.

\paragraph{Bayesian inference.}\label{sec:mcmc}

\begin{figure}
    \centering
    \includegraphics[width=0.70\textwidth]{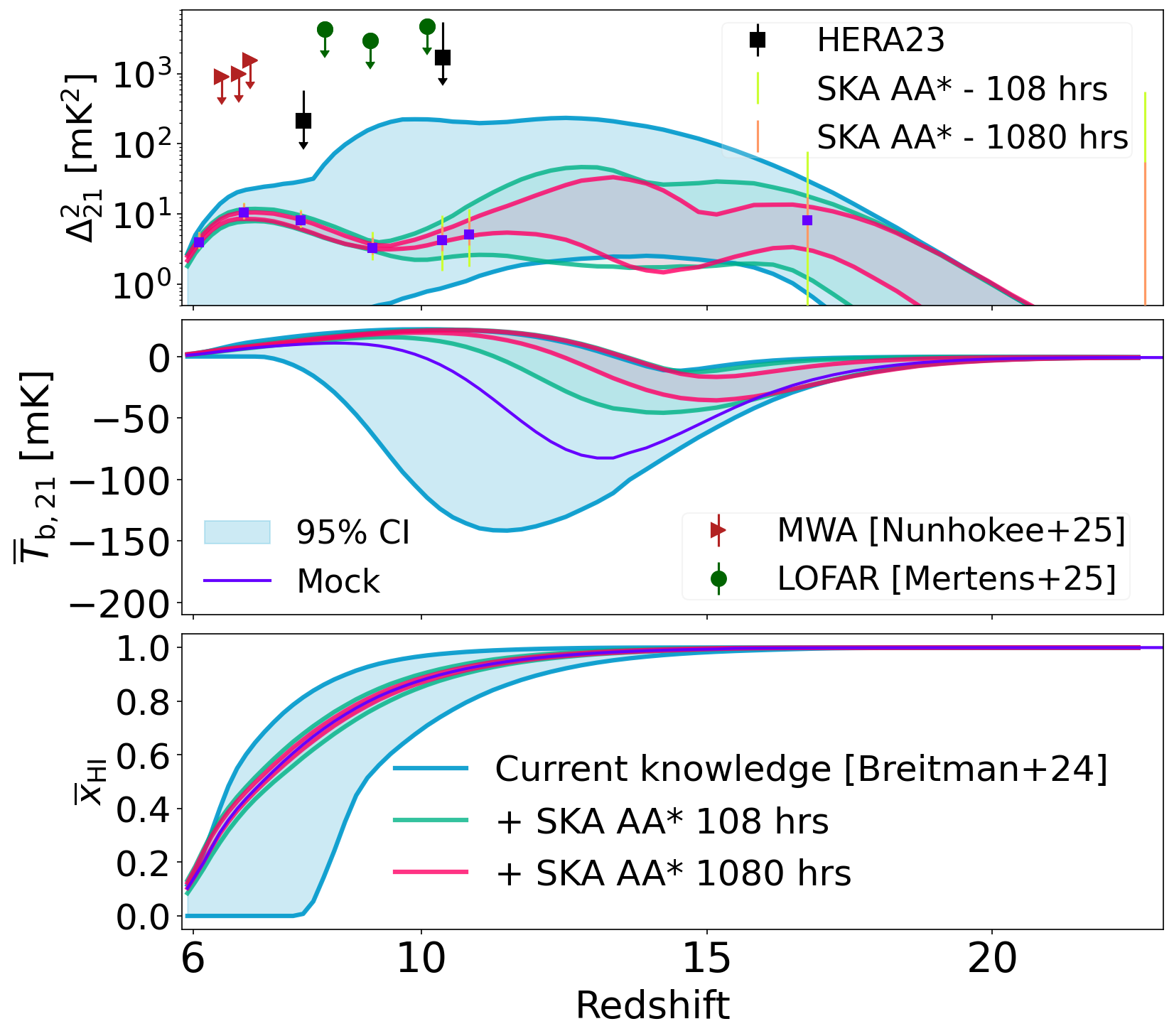}
    \caption{Posteriors from three different inferences for the 21\,cm power spectrum (top panel), the 21\,cm global signal (middle panel), and the EoR history (bottom panel). The blue posterior shows current state-of-the-art inference results \citep{BreitmanMesinger_2024} obtained with the {{\sc 21\,cmEMUv1}} emulator and a combination of four observables, including the to-this-day lowest upper limits on the 21\,cm 1D PS (black squares; \citealt{Hera2023}). The results are compared with posteriors obtained for 1080hrs (magenta) and 108hrs (green) of SKA AA$^\ast$ observations alone.}
    \label{fig:inference_results}
\end{figure}

Bayesian inference is a statistical method to robustly derive the probability of a model $\mathcal{M}(\theta)$ with parameters $\theta$, given a set of observations $d$ and \textit{priors} $P(\theta)$ about the parameters. This prior can be anything: it can be deliberately ignorant or it can be informative, for example, by reflecting previous observations. Given observations $d$, we update these priors into a \textit{posterior} $P(\theta | d)$, incorporating the information from the data $d$ via Bayes' theorem: 
\begin{equation}
\label{eq:bayes}
    P(\theta | d) = \frac{P(d|\theta) P(\theta)}{P(d)},
\end{equation}
where $P(d|\theta)$ is the \textit{likelihood} of measuring $d$ given model parameters $\theta$, and $P(d) = \int \rm{d}\theta \ P(d|\theta) P(\theta)$ is the model \textit{evidence}.
In this section, we illustrate the power of SKA-Low to infer properties of the CD and EoR by performing a set of inferences with simulated mock 21\,cm 1D PS observations, obtained with EOS21 \citep{Munoz22} and shown as purple squares 
in the top panel of Fig. \ref{fig:inference_results}. We simulate the SKA-Low sensitivity with {\sc 21\,cmSense} \citep{Pober13, Pober14, Murray24}, 
for two surveys with the AA$^\ast$ layout: 180 days (1080 hrs) and 18 days (108 hrs), observing a single sky patch for 6 hours each day\footnote{Since the telescope always tracks the same patch of sky, the sensitivity at large scales is sample-variance limited. A less deep survey observing multiple fields would reduce sample variance and improve sensitivity on large scales.}. For comparison, in Fig. \ref{fig:inference_results} we also show current deepest upper limits from MWA \citep{TrottNunhokee_2025}, LOFAR \citep{MertensMevius_2025}, and HERA \citep{Hera2023} in red, green, and black, respectively. 
We assume a `moderate' foreground scenario, where we exclude all $k$-modes below the horizon limit plus a buffer of $\sim 0.07$ Mpc$^{-1}$.

We use {\sc 21\,cmEMUv1} \citep{BreitmanMesinger_2024} as the forward model and we assume a standard Gaussian likelihood on the 21\,cm 1D PS and a uniform prior for all parameters (see \citealt{BreitmanMesinger_2024} for the exact prior ranges) in Eq.~\eqref{eq:bayes}. We perform Bayesian inference using {\sc 21\,cmMC} \citep{GreigMesinger_2015} with the {\sc MultiNest} sampler \citep{Feroz09}, which implements a nested sampling algorithm\footnote{Several other algorithms exist to sample the posterior distribution. These include Markov Chain Monte Carlo (MCMC) methods such as Metropolis-Hastings \citep{Metropolis_1953,Hastings_1970}, and Hamiltonian Monte Carlo \citep[HMC;][]{JascheKitaura_2010,BetancourtGirolami_2015}, which can achieve faster convergence in high-dimensional parameter spaces. Nested sampling offers the advantage of calculating the Bayesian evidence, which is crucial for model comparison \citep[e.g.,][]{BinniePritchard_2019}.} \citep[see, e.g.,][for a review]{Skilling_2004}. The analysis requires about 120\,000 emulator evaluations to converge. The 95\% credible intervals (CI) of the posterior distributions are shown in Fig.~\ref{fig:inference_results} in magenta (green) for 1080 hrs (108 hrs), demonstrating the improvement brought by the SKA alone compared to current knowledge (blue), based on four observables (see \citealt{BreitmanMesinger_2024} for more details): (i) the HERA upper limits \citep{Hera2023}; (ii) UV luminosity functions \citep{Bouwens15, Bouwens16, Oesch18}; (iii) the Thomson scattering optical depth to the CMB \citep{Planck18}; and (iv) an upper limit on the neutral fraction at $z = 5.9$ \citep{McGreer15}.

Fig.~\ref{fig:inference_results} illustrates that only 108hrs of SKA AA$^\ast$ observations are sufficient to constrain the IGM neutral fraction down to $\sim$ percent-level: the EoR midpoint is recovered to within $\pm 0.02\ (0.04)$ for 1080hrs (108hrs) at 95\% CI. We constrain the Thomson scattering optical depth 68\% CI over 22 (7.5) times tighter than current observations (e.g., \citealt{Qin20, IlicTristram_2025}) with 1080hrs (108hrs). The 21\,cm 1D PS alone, however, is not sensitive enough to constrain all model parameters and the global signal (middle panel) is not accurately recovered. 
These results rely on the simplifying assumption that the 21\,cm PS follows a Gaussian distribution, supporting the use of a Gaussian likelihood. In reality, the 21\,cm signal arises from complex, highly non-linear astrophysical processes, and even its PS exhibits non-Gaussian features. Consequently, the true likelihood of the 21\,cm PS is analytically intractable. To overcome this limitation, simulation-based inference methods have been proposed, which is discussed next.

\paragraph{Simulation-based inference.}

While classical inference requires an explicit likelihood, which is rarely (if ever) available with summary statistics of the 21\,cm signal, Simulation-Based Inference (SBI) is a family of methods that bypasses this requirement. A set of simulations of the 21\,cm signal (affected by the relevant instrumental processes) is a set of draws in the joint distribution of the data and the parameters. It can be modelled by a flexible parametric function using various machine-learning methods, fitting either the implicit likelihood of the model, the ratio of the likelihood and the evidence, or the posterior distribution. SBI has received significant attention in the 21\,cm community \citep[see, e.g.,][]{Neutsch:2022, Prelogovic23,  Heneka:2023,SaxenaCole_2023,Schosser:2024,Meriot24,Meriot24b, Zhao22,Zhao24,Pietschke:2025}, and can in principle be used with any summary statistics or combination of summaries \citep{Meriot24b, Prelogovic24, Schosser:2025}, fixed or learned (see Sec.~\ref{sec:info_max}).

In addition to benefitting from the true likelihood of the modelling pipeline and typically requiring fewer samples in the training set than steps in a single MCMC run, SBI methods have the desirable trait of amortising the computation cost\footnote{In Sec. \ref{sec:mcmc}, we discuss an alternate method to reduce this cost, which is by replacing simulation code with emulators.}. Classical MCMC chains are (at least partly) sequential, leading to poor scaling of the computation cost with the number of inferences. However, the cost of SBI often lies `upfront', in the generation of the simulation set and the training of the networks, which dwarfs the cost of drawing from the posterior after the training. This allows cheap statistical validations of inference pipelines on sets of posteriors \citep[see, e.g.,][]{Prelogovic23,Meriot24,Schosser:2024,Schosser:2025,CerardiGiri_2025}, as well as the use of relatively expensive simulation codes. 
\citet{Prelogovic23} found that the 1D PS likelihood is well described by a Gaussian, as long as (i) it includes the covariance between wave-modes and redshift bins and (ii) this covariance is estimated from an average across parameter space, and not from a single realisation or at a single point in parameter space. Such shortcuts tend to bias and over-constrain posteriors. By contrast, SBI learn the likelihood implicitly, capturing non-Gaussian features and parameter-dependent covariances and delivering accurate posteriors with up to ten times fewer simulations than explicit-likelihood MCMC approaches.

\paragraph{Direct parameter prediction.} 
A different type of analysis allows one to infer parameter values directly, with no posterior distribution of possible values and no error estimation.
Early studies that adopted feed-forward multilayer perceptrons (MLP) type artificial neural networks (ANNs) used for point estimation (i.e. predicting parameters without error bars or posterior) showed that low‑dimensional summaries of the 21\,cm signal already carry enough information for reliable astrophysical inference. \citet{shimabukuro2017analysing} demonstrated that a basic MLP fed with 21\,cm 1D PS at several redshifts can reconstruct reionisation parameters with high fidelity, and \citet{Doussot2019} refined this idea, comparing MLPs with kernel‑based regressors and reporting clear accuracy gains. \citet{2022Choudhury} proposed a two-stage framework in which one MLP extracts the 21\,cm 1D PS from foreground-contaminated observations and a second MLP converts that 21\,cm 1D PS into physical parameters. Beyond two‑point statistics, \citet{JenningsWatkinson_2020} employed MLP on three‑point correlation functions and confirmed that higher‑order information improves the reconstruction of bubble and ionisation metrics. A parallel thread targets the 21\,cm global signal. \citet{choudhury2020extracting} showed that an MLP can jointly disentangle foregrounds and astrophysical parameters, while a follow‑up by \citet{choudhury2021using} applied the network to EDGES data, illustrating practical applicability. \citet{Tripathi2024} extends this line by including ionospheric distortions and still obtains the desired parameters robustly. These works highlight the flexibility of simple networks for parameter estimation from 21\,cm statistics.

More recent works have applied convolutional neural networks (CNNs) to exploit the full spatial structure of simulated observations. For example, \citet{Gillet2019} trained a 2D CNN on slices of 21\,cm tomographic cubes and retrieved multiple source properties with accuracy that rivals traditional Bayesian inference pipelines, and \citet{HassanLiu_2019} found that these CNN can be employed to identify the sources of reionisation. 
Pushing to fully volumetric data, \citet{Prelogovic2022} compared a 3D RNN against a CNN, both of which ingest light‑cone cubes and predict astrophysical parameters while remaining resilient to foreground residuals.
\citet{Neutsch:2022} extended the 3D CNN architecture to simultaneously infer astrophysical and dark‑matter parameters. Together, these CNN‑based studies underline the value of deep architectures that ingest images or volumes directly, capturing non‑Gaussian information that escapes lower‑order summaries.
Furthermore, several methods have been developed to directly infer the ionisation history by using the patterns in the 21\,cm maps. For example, \citet{MangenaHassan_2020} used CNN to map the patterns in the 21\,cm maps to the ionisation fraction, while \citet{BiancoGiri_2021,BiancoGiri_2024} trained a U-Net to identify patterns in these maps. Studies such as \citet{GiriMellema_2018b} developed model-independent methods to identify structures and estimate the ionisation history. We refer the interested readers to \cite{Acharya02.2026.SKA} and \citet{Bag01.2026.SKA} for more discussion.

\section{Information content of statistics beyond the power spectrum} \label{sec:beyond_PS}

In this section, we examine the amount of astrophysical information contained in summary statistics other than the power spectrum. In Sec.~\ref{sec:higher_order}, we introduce statistics of order more than two (`higher-order statistics') and image-based statistics in Secs.~\ref{subsec:moments}, \ref{sec:bubble_stats} and \ref{sec:time_fields}. In Sec.~\ref{sec:Fisher_Discussion}, we compare their information content through their Fisher information matrices.

\subsection{Statistics beyond order two}
\label{sec:higher_order}

The redshifted 21\,cm radiation from the EoR is expected to be highly non-Gaussian. Indeed, the zero pixels coming from ionised regions will lead
to a bimodal (and not normal) distribution of pixel temperatures. Therefore, non-Gaussianities are linked to the presence of ionised regions in a map and can be used to characterise reionisation and the nature of ionising sources, the state of the IGM and the physical processes within the IGM \citep{Bharadwaj:2004sx, IlievMellema_2006, MellemaIliev_2006}. Statistics of order higher than two, beyond the power spectrum, commonly called `higher order statistics', inherently measure this non-Gaussianity.

\subsubsection{Bispectrum and its derivatives} 
\label{sec:bispectrum}

\paragraph{The 21\,cm bispectrum.} One way to quantify the non-Gaussianity in the signal is the bispectrum  $\mathcal{B}(\bm{k_1},\bm{k_2},\bm{k_3})$ of the 21\,cm  differential brightness temperature field $\delta T_{b}(\bm{r})$, defined as 
\begin{align} \label{eq:def_bispec}
    \langle \tilde{T}_{\rm b}(\bm{k}_1) \tilde{T}_{\rm b}(\bm{k}_2) \tilde{T}_{\rm b}(\bm{k}_3) \rangle =V\, \delta^D(\bm{k}_1+\bm{k}_2+ \bm{k}_3)~\mathcal{B}(\bm{k}_1,\bm{k}_2,\bm{k}_3),
\end{align}
where $\tilde{T}_{\rm b}(\bm{k})$ is the 3D Fourier transform of $\delta T_{b}(\bm{r})$ and $V$ is the observed volume. The Kronecker delta function $\delta^D(\bm{k}_1+\bm{k}_2+ \bm{k}_3)$ ensures that only closed triangles in Fourier space contribute. 
The 21\,cm bispectrum is a real quantity even though estimated from the product of three complex visibilities \citep{ Majumdar_bispec2018}. It measures the excess probability between three points forming a triangle in Fourier space and, as such, will be sensitive to the morphology and distribution of ionised regions during the EoR \citep{HutterWatkinson_2020, Noble_bispec2024} and to the distribution of neutral islands at its end stages \citep{RasteKulkarni_2024, GillPramanick_2024}. Different unique triangle shapes formed by the wave-vectors will be sensitive to the correlation between the signal at different length scales. For example, squeezed triangles ($k_3 \ll k_1, k_2$) contain the correlation between the small and large scales \citep{LewisChallinor_2011}.
The evolution in the bispectrum sign and amplitude with the progress of reionisation can reveal the morphological transformation of the 21\,cm field \citep{WatkinsonGiri_2019, Majumdar_bispec2018, 2020_Majumdar, HutterWatkinson_2020, GillPramanick_2024, Noble_bispec2024, TiwariShaw_2022, KamranGhara_2022,2021MNRAS_Kamran}. In particular, it is possible to link the scale at which the bispectrum flips sign with the size of the ionised bubbles (Sec.~\ref {sec:bubble_stats}).

\textit{Constraining power.} Several works have demonstrated that the bispectrum can independently put tighter constraints on the CD-EoR parameters than the 1D PS (Sec.~\ref{sec:PS}). Further, combining bispectrum and PS measurements for inference breaks degeneracies on certain EoR model parameters such as the virial temperature of halos or the ionising efficiency of light sources \citep{2016:ShimabukuroYoshiura,ShimabukuroYoshiura_2017, WatkinsonGreig_2022, KamranGhara_2022,2021MNRAS_Kamran,TiwariShaw_2022,Noble_bispec2024,Mahida_2025}.
In Sec.~\ref{sec:Fisher_Discussion}, we employ the bispectrum, in the squeezed-limit, within a Fisher analysis framework to assess its information content on key astrophysical parameters. The binned 21-cm bispectrum, in the squeezed-limit, is estimated using the fast Fourier transform–based estimator of \citet{2021:ShawBharadwajSarkar}, which follows the unique triangle-shape classification of \citet{Bharadwaj+2020}. The analysis uses the full covariance matrix; however, we find that off-diagonal contributions are negligible for the present setup. This behaviour may vary with differences in the simulation techniques used to generate the 21-cm signal, as noted by \citet{Mondal_2017}.

\textit{Observability.} One can build a bispectrum estimator from observations as a correlation between three visibilities \citep{Bharadwaj:2004sx,Yoshiura2015}.
\citet{TrottWatkinson_2019} found that some wave-modes of the 21\,cm bispectrum might be detectable before the power spectrum, in under 100 hours of observations with the MWA. However, \citet{WatkinsonTrott_2021} showed that measurements can be strongly corrupted by instrumental systematics and dominated by foreground contamination on all scales considered. A multi-frequency angular bispectrum estimator can instead separate the complex foreground wedge from the cosmological signal, leading to the first upper limits on the bispectrum, obtained with MWA observations \citep{2019PASA...36...23T,2025_Gill}.  

\paragraph{The triangle correlation function of phases.}
Because it is defined in Fourier space and is a function of two Fourier modes, the bispectrum $\mathcal{B}(\bm{k}, \bm{q})$ is difficult to interpret. Another possibility is to define statistics based on the bispectrum, but defined in real space \citep{JenningsWatkinson_2020}. This is the idea behind the triangle correlation function of phases $s(r)$ \citep[TCF,][]{GorcePritchard_2019}, which measures the probability of an equilateral triangle of side length $r$ to be located within an ionised bubble in the 21\,cm signal. A TCF estimator can be computed numerically from a simulation of side length $L$ in $D$ dimensions:
\begin{equation}
s(r)=\left(\frac{r}{L}\right)^{3 D / 2} \sum_{k, q \leq \pi / r} \omega_D(p r) \frac{\mathcal{B}(\bm{k}, \bm{q})}{|\mathcal{B}(\bm{k}, \bm{q})|} .
\end{equation}
For a three-dimensional simulation box, $\omega_3(x)$ is the Bessel function of the first kind and order zero. The $p$ vector is a function of $r$ and performs the selection of equilateral triangles in real space. The TCF is a tracer of the ionisation level and has been shown to recover the (maximum) bubble size on simulated ionisation fields of various complexities. In this chapter, we test for the first time its performance in constraining astrophysical parameters (Sec.~\ref{sec:Fisher_Discussion}).

\subsubsection{Position-dependent power spectrum}\label{subsec:posdep_ps}

Unlike the standard power spectrum, which quantifies average fluctuations at a given scale, the Position-Dependent Power Spectrum (PdPS) measures how the small-scale power spectrum is modulated by the large-scale environment \citep{ChiangWagner_2014}. This statistic provides a computationally efficient way to access non-Gaussian information encoded in the squeezed-limit bispectrum without requiring the direct calculation of three-point correlations. It quantifies how small-scale 21\,cm power responds to the large-scale environment--particularly relevant during reionisation, where large ionised bubbles are modulated by smaller-scale clustering of ionising photon sources. 
\citet{GiriDAloisio_2019} first applied it to the 21\,cm signal and showed it can be measured with SKA-Low. 

The PdPS is derived by dividing a volume into sub-regions centred at $\pmb{r}_L$, calculating their local power spectrum $P(k,\pmb{r}_L)$, and correlating it with the mean signal inside the sub-regions $\overline{\delta}(\pmb{r}_L)$, which quantifies large-scale fluctuations of the signal\footnote{The mean 21\,cm signal inside sub-regions can be non-zero while the mean of the full volume is zero, which is true for any radio interferometric data due to absence of zero baseline \citep{GiriDAloisio_2019}. However, see Sec.~\ref{subsec:3_moments_mean} for more discussion about advanced methods explored to measure the mean signal from interferometers.}. The correlation between the local power spectrum and the large-scale mean field writes \citep{ChiangWagner_2014}:
\begin{equation}
    \label{eq:pdps_bispectrum}
    \langle P(k,\pmb{r}_L) \overline{\delta}(\pmb{r}_L) \rangle = \frac{1}{V_{L}^{2}}\int\frac{d^{3}q_{1}}{(2\pi)^{3}}\int\frac{d^{3}q_{3}}{(2\pi)^{3}}\mathcal{B}(k-q_{1},-k+q_{1}+q_{3},-q_{3}) W_{L}(q_{1})W_{L}(q_{1}+q_{3})W_{L}(q_{3})
\end{equation}
where $V_L$ is the sub-volume, $W_L$ the window function defining it, and $\mathcal{B}$ the bispectrum (Sec.~\ref{sec:bispectrum}). This integral is dominated by squeezed configurations $q_1\approx q_2 \gg q_3$. The PdPS is often normalised through a \textit{response function} $f(k) \equiv \langle P(k,\pmb{r}_L) \overline{\delta}(\pmb{r}_L) \rangle/(\sigma^2_L P(k))$, where $\sigma^2_L$ is the variance of $\overline{\delta}(\pmb{r}_L)$ within sub-regions. For the 21\,cm PdPS, two distinct response functions can be constructed: One correlating small-scale 21\,cm power with large-scale 21\,cm fluctuations (auto-squeezed bispectrum), the other with large-scale dark matter fluctuations (cross-squeezed bispectrum).

\citet{GiriDAloisio_2019} showed the response function exhibits distinctive redshift evolution sensitive to reionisation topology. For example, in inside-out scenarios where dense regions ionise first, the correlation between small-scale 21\,cm power and large-scale density flips from positive to negative as ionisation progresses, with markedly different evolution for outside-in scenarios. Because the PdPS measures a specific mode-coupling signature, it may be harder for foregrounds or systematics to mimic compared to the power spectrum alone, offering potential for signal validation. SKA-Low's ability to measure the PdPS across redshifts will provide crucial constraints on reionisation topology, complementing the power spectrum and other higher-order statistics.

\subsubsection{Scattering Transforms} 
\label{sec:scattering}

Scattering Transforms have emerged as a powerful summary statistic for characterising non-Gaussian fields \citep{2011:Mallat, 2012:BrunaMallat}. Inspired by convolutional neural networks, scattering transforms are constructed through successive wavelet convolutions and the application of non-linear operators, such as modulus probing fields, at increasingly fine spatial scales and orientations \citep{2019:AllysLevrierZhang, 2020:ChengTingMenard}.
Scattering transform-based statistics, including wavelet scattering transforms (WST), wavelet phase harmonics (WPH) and, in particular, scattering covariances, have demonstrated robust and reliable performance in capturing complex non-Gaussian features \citep{2019:AllysLevrierZhang, 2020:ChengTingMenard, 2020:AllysMarchandCardoso, 2023:ChengMorelAllys}. Compared to traditional $n>2$ point statistics, such as the bispectrum, these methods offer significantly reduced estimator variance, improved stability, and enhanced interpretability \citep{2019:AllysLevrierZhang,2023:ChengMorelAllys}, motivating their usage for astrophysical and cosmological parameter inference, especially in high-redshift 21\,cm cosmology \citep{2022:GreigTingKaurov,HothiAllys_2024,Prelogovic24,2024:ZhaoMaoZuo,2025:ShimabukuroXuShao,2025:HothiAllysSemelin}. A more detailed overview of the Scattering Transform can be found in the \citet{Bag01.2026.SKA}.
In Sec.~\ref{sec:Fisher_Discussion}, we focus on the 2+1 (`evolution-compressed') WST statistic of \citet{HothiAllys_2024} for EoR LC analysis. The 2D WST is first applied to each frequency slice, yielding coefficients that capture spatial information at fixed redshift. The evolution of each coefficient along the LC (across frequency/redshift) is then summarised by a 1D continuous wavelet transform with a dyadic cosine wavelet and compressed via integrated $\ell_1$- and $\ell_2$-norms across scales.

\subsection{Moments} \label{subsec:moments}

To extract physical information from 21\,cm data, one can also study the pixel probability distribution function (PPDF) of a 21\,cm intensity map and its moments \citep{CiardiMadau_2003,furlanetto2004growth, MellemaIliev_2006}, driven by the morphology of the ionised regions during the EoR. For example, for a binary ionisation field, where pixels are either fully ionised or fully neutral, the PPDF can be derived as a combination of Dirac delta functions $\delta$:
\begin{equation}
\label{eq:ppdf}
    P(x) = (1-\bar{x})\, \delta (x) + \bar{x}\, \delta (x-1),
\end{equation}
where $\bar{x}$ is the expectation value of the distribution, or the global 21\,cm signal (Sec.~\ref{subsec:3_moments_mean}). Below, we describe what physical information can be extracted from the statistical moments of the PPDF.

\paragraph{Mean: The global signal.} \label{subsec:3_moments_mean}
The global signal traces the sky-averaged brightness temperature as a function of redshift, and encodes the broad evolution of the signal across time. Typically, it is measured with a single receiving element, which collects total power; interferometers, which measure fluctuations on given spatial scales, are generally not sensitive to the total power signal, except in their auto-correlations. Several works in the literature have looked at the possibility of measuring the global signal with interferometric experiments \citep[e.g.,][]{thekkeppattu_sitara_global,mckinley2018measuring,ZhangXin_2023,IgnatovPritchard_2024}, with measurements undertaken by SITARA that demonstrated the ability to extract the zeroth-order moment from the shortest spacings. This short spacing mode is one avenue for extracting the global signal with internal SKA-Low station dipoles, enabled by their individual digitisation. 
The interferometric approach has advantages over the single element approach, which requires a precision sky-based calibration, or well-characterised external noise source for instrument calibration. Interferometers avoid this issue somewhat by the additional measurements in the system. 
The EDGES experiment, which claimed a detection in 2018 \citep{edges}, undertook extensive efforts to separate the chromatic beam response to the sky signal. A following work \citet{saras3} used SARAS3 \citep{Nambissan_2021}, a different antenna and backend system, to rule out the EDGES signal as being purely cosmological at a $2\sigma$ level, highlighting the difficulty with this experiment. On the other hand, one benefit of the SKA-Low system is its very wide instantaneous bandwidth, which allows for better characterisation of Galactic emission. With AA* and AA4's collecting area, the sensitivity to the global signal may be large. However, both the mutual coupling between antennas, which introduces direction-dependent phase, and the need for a precise frequency-dependent understanding of the primary beam, may make this challenging. 

\paragraph{Variance.}

From the analytical probability distribution function in equation~\eqref{eq:ppdf}, all the statistical moments of the distribution for a binary field can be derived. Comparing how the statistical moments derived numerically from more sophisticated models and simulations deviate from the analytical expression provides us with hints on the nature of reionisation \citep{WyitheMorales_2007,Gluscevic_Barkana_2010, IchikawaBarkana_2010}, such as its reionisation topology \citep[e.g., inside-out or outside-in, see][]{WatkinsonPritchard_2014}, its global reionisation history \citep{BittnerLoeb_2011,Patil_2014}, and galaxy properties \citep{KubotaYoshiura_2016}, even when derived from dirty 21\,cm signal images or after foreground removal \citep{HarkerZaroubi_2009, Patil_2014, KittiwisitBowman_2018}. However, these one-point statistics lack information on the correlations between pixels. For this reason, \citet{BarkanaLoeb_2008} have extended this formalism by analysing the one-point PDF of the difference in the differential 21\,cm brightness temperature measured at two points and \citet{GorceHutter_2021} have introduced the local variance, which describes the distribution of the mean differential 21\,cm brightness temperatures measured in two-dimensional maps along the LC. \citet{2019MNRAS.486.5766T} used a kernel density estimation approach to MWA data to separate Gaussian from non-Gaussian components, exploring the utility of such an approach to treating foreground contamination.

\paragraph{Skewness \& Kurtosis.}

The skewness of the one-point statistics of the brightness temperature distribution is the third central moment normalised by the variance. Skewness flips sign and amplitude in concert with key astrophysical transitions -- remaining near zero during the Gaussian-dominated dark ages, becoming negative as Lyman-$\alpha$ coupling drives the spin temperature below the CMB, and then rising to a positive maximum once X‑ray heating becomes effective \citep{ShimabukuroYoshiura2015}. 
\citet{HarkerZaroubi_2009} showed that skewness exhibits a dip as nascent ionised bubbles first appear—producing a low‑temperature tail—and then a steep rise as ionised regions percolate, a pattern recoverable even after foreground mitigation and with SKA precursors \citep[see also][]{KittiwisitBowman_2022}. \citet{watkinson2015impact} further quantified how variations in X‑ray efficiency and spectral hardness shift the redshift of the skewness minimum and maximum, providing constraints on high‑redshift source properties.
Some recent works have moved these analyses from simulations to (mock) observations and estimated the sensitivity of current and future interferometers to the $n$th moments: \citet{CookBalu_2024} have characterised the cosmic‑variance limits on these non‑Gaussian estimators and \citet{KittiwisitBowman_2018} demonstrate that, accounting for systematics from instrument configuration, thermal noise, sample variance, and the LC effect, the variance, skewness, and kurtosis can be measured with the full HERA array. More recently, \citet{MaPeng2023} adapted such statistics to Fourier space -- defining  a skew spectrum and smoothed skewness that mirror bispectrum features but offer substantially higher signal‑to‑noise ratios for SKA-Low.

\subsection{Bubble statistics} \label{sec:bubble_stats}

Characterising the morphology of ionised regions during reionisation is essential for understanding this epoch. While the 21\,cm signal is sensitive to the distribution of neutral hydrogen fraction ($x_\mathrm{HI}$), this information is mixed with cosmological information, such as matter and temperature fluctuation (see Eq.~\ref{eq:dTb}), and contaminants, such as instrumental noise and foreground residuals \citep{BiancoGiri_2024}. Therefore, extracting information about the bubbles depends on our ability to identify them in 21\,cm tomographic data, which will become feasible with SKA-Low \citep[e.g.,][]{GharaChoudhury_2017,GiriMellema_2018b}. Several techniques have been developed to extract \hii{} regions from noisy 21\,cm measurements: \citet{GiriMellema_2018b} developed a method based on image over-segmentation (superpixels) that optimally identifies boundaries of physical patterns in noisy data. More advanced approaches using neural networks, particularly U-net architectures, have further improved this capability \citep[e.g.,][]{GagnonHartmanYue_2021,BiancoGiri_2021,BiancoGiri_2024,BiancoGiri_2025}.
We refer the interested reader to \citet{Bag01.2026.SKA} for more details. 

\paragraph{Bubble sizes.}

The growth and distribution of ionised bubbles during the epoch of reionisation is a direct probe of the nature of the first ionising sources \citep{FurlanettoMcQuinn_2006}. A useful metric to study their statistical properties is to construct a bubble size distribution (BSD, see also Sec.~4.1 of \citet{Bag01.2026.SKA}).
The BSD has been examined in a multiplicity of methods, from theoretical constructions \citep[e.g.,][]{furlanetto2004growth,DoussotSemelin_2022} to sampling the ionisation field in simulations \citep[e.g.,][]{LinOh_2016,GiriMellema_2018a}. A non-exhaustive list of examples of numerical methods for the latter include the friends-of-friends algorithm \citep{IlievMellema_2006}, the spherical average method \citep{ZahnLidz_2007}, the mean free path method \citep{21cmFAST_2007}, granulometry \citep{KakiichiMajumdar_2017}, the watershed \citep{LinOh_2016}. Note that the latter methods can be applied to the 21\,cm signal data after identifying the ionisation field using frameworks mentioned above. 

\paragraph{Topological measures.}\label{sec:topo_measures}

Topological measures are designed to quantify the morphology (shape and geometry) and topology (connectivity) of the complex network of ionised bubbles in 21\,cm brightness temperature LCs.
Historically, the most widely used topological descriptors have been the Minkowski Functionals (MFs): in 3D, the volume, the surface area, the mean integrated curvature, and the Euler characteristic ($\chi$). The Euler characteristic, and its equivalent the genus, quantifies the connectivity of ionised regions, distinguishes different reionisation scenarios, and identifies distinct phases of the EoR \citep{2006:Gleser, 2008:Lee, FriedrichMellema_2011, 2014:Hong, 2015:Wang, YoshiuraShimabukuro2017, 2019:chen}. With the tomographic 3D images of the EoR provided by SKA-Low, direct measurement of the Euler characteristic will be possible \citep{GiriMellema_2019}, and combining the 3D MFs with the power spectrum is forecasted to significantly improve constraints on reionisation model parameters \citep{2024:Diao}. A further generalisation, the Minkowski Tensors (MTs), can capture anisotropic information, providing details on the shape and orientation of structures beyond the scalar MFs. 
In Sec.~\ref{sec:Fisher_Discussion}, we quantify the information content of the Euler characteristic, through its Fisher information matrix. We refer the reader to the \citet{Bag01.2026.SKA} for a description of additional topological measures such as the Betti numbers, largest cluster statistics, and persistent homology.

\subsection{Reionisation Time Fields}
\label{sec:time_fields}

The EoR can be described in terms of the reionisation time field, $\TR$ or, equivalently, the reionisation redshift field $\ZR$, which represents a distribution of times when each point in space $\Vec{r}(x, y, z)$ became reionised \citep{Trac2007}. A map of the reionisation redshift can indeed be used to reconstruct the entire reionisation history, morphology included, of the corresponding region of the sky and, therefore, to fully characterise reionisation and its sources \citep{Thelie2022}.
As such, $\ZR$ can be been used as the basis for a reionisation model: \citet{Battaglia2013} have parameterised its power spectrum to quickly generate $\ZR$ simulation boxes, whilst \cite{Trac2022} introduced AMBER, a semi-analytical model of the EoR which assigns a redshift of reionisation to halos based on their mass and formation history. 
Building on the observation that the $\TR$ field is close to Gaussian, \cite{Thelie2023} demonstrated that its statistical properties, such as its PPDF, can be predicted theoretically by validating their approach on numerical simulations. The Gaussian property of $\TR$ is crucial because it allows to predict the reionisation process from only its power spectrum, opening a path towards an analytical model of the EoR. \cite{Chardin2019} used convolutional neural networks (CNNs) to predict the $\TR$ field from the source density and gas density fields, providing a computationally efficient method to describe the entire reionisation process without explicitly modelling all underlying physics. Analysing simulations of the local Universe with $\ZR$ have revealed that reionisation is not uniform across different galactic environments: \cite{Ocvirk2014}, \cite{Aubert2018}, and \cite{Sorce2022} found that galaxies in the Local Group experience their own distinct reionisation timelines, highlighting the complex and inhomogeneous nature of the process.
On larger scales, using $\ZR$ maps, \cite{Trac2008} explored the global thermal imprint of  reionisation sources and \citet{Deparis2019,zhu2025} measured the velocity of ionisation fronts. 
Despite this theoretical potential, $\TR$ is not observable directly. \citet{Hiegel2023} have explored the potential of CNNs to reconstruct $\TR$ maps from 21\,cm maps measured by SKA-Low near the midpoint of reionisation. 
Such reconstructions can then be used to exclude extreme reionisation models, e.g., based on Warm Dark Matter \citep{Hiegel2024}.

\subsection{Discussion of the Fisher analysis}
\label{sec:Fisher_Discussion}

In Sec.~\ref{sec:PS_fisher}, we showed that the 21\,cm power spectrum can tightly constrain astrophysical parameters during the EoR, with precision improving for longer integrations and with the AA4 layout. While the first detection of the 21\,cm signal will likely come from its spherically averaged power spectrum, image-based and beyond-power-spectrum statistics provide additional information. Using the Fisher analysis outlined in Appendix~\ref{sec:fisher_framework}, we quantify the information each statistic probes in the AA* noise regime; the full results and analysis codes are publicly available\footnote{See \url{https://github.com/sambit-giri/ska_eor_inference}.}. The Fisher matrix assumes a multivariate Gaussian likelihood \citep{1997:TegmarkTaylorHeavens}, an approximation that is reasonable for two-point statistics such as the power spectrum but can break down for strongly non-Gaussian descriptors (e.g. the bispectrum, Minkowski functionals, or topological measures) \citep{2013:Carron}. In those cases, the Fisher formalism may underestimate parameter degeneracies and overstate precision, but it remains a useful first-order diagnostic of the \emph{information content} and \emph{parameter sensitivity} of different observables.

Across all statistics, increasing the integration time from 100 to 1000 hours boosts the Fisher information, summarised by the Figure of Merit (FoM), defined as the square root of the determinant of the Fisher matrix and proportional to the inverse of the three-dimensional posterior volume in parameter space. For the AA* configuration, Table~\ref{tab:FoM_AAstar} shows that the FoM rises with integration time for most observables, as expected from reduced thermal noise and improved sampling of the 21\,cm signal. In particular, the FoM increases by a factor of 588 for the bispectrum and 147 for PS1D. These results confirm that deeper SKA-Low observations enhance the constraining power of all summary statistics, with each probing complementary aspects of the signal.
Higher-order statistics, such as the bispectrum, are especially sensitive to non-Gaussian features of the 21\,cm field and therefore probe information beyond that contained in two-point statistics. In our forecasts, the bispectrum carries more Fisher information on the astrophysical parameters in the squeezed limit than the power spectrum, underscoring the importance of characterising non-Gaussianity for understanding the EoR.

\begin{table}[h]
    \centering
    \begin{tabular}{l|cc|cc|ccc}
        & Moments & Topology & WST & Bispectrum & PS1D & PD2D & MAPS \\
        \hline
        \hline
        Factor improvement & 1.32 & 1.29 & 19.7 & 588 & 147 & 10.2 & 41.0 \\
    \end{tabular}
    \caption{Improvement in the figure of merit on three astrophysical parameters, moving from 100h to 1000h integration time with SKA-Low AA*, for various statistics.}
    \label{tab:FoM_AAstar}
\end{table}

\begin{figure}
    \centering
    \includegraphics[width=0.65\linewidth]{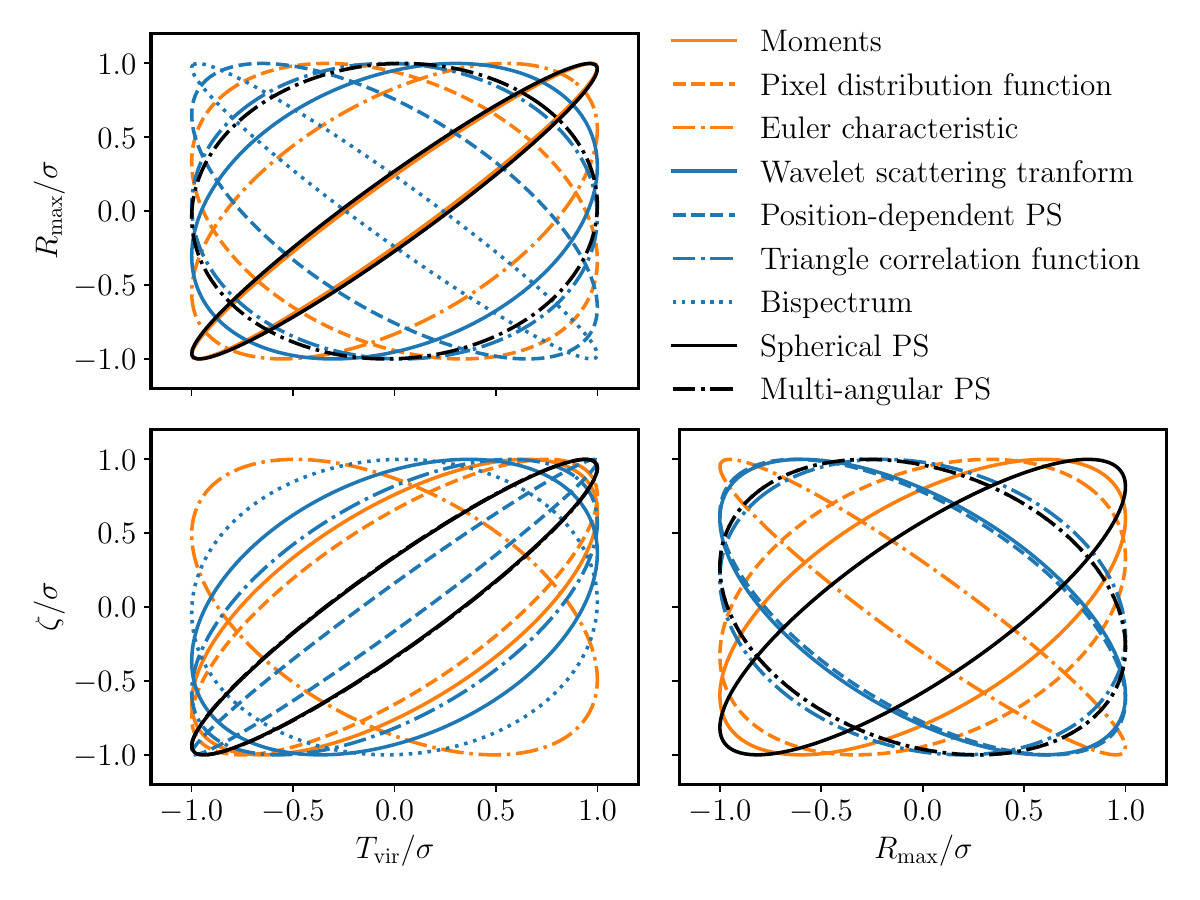}
    \caption{Normalised $1\sigma$ contours for 1000 hours of observations with the AA* configuration, from a Fisher forecast based on different summary statistics. Because each statistic probes different physical information, the degeneracy lines vary. Combining power spectrum measurements with other statistical measurements will, therefore, allow to break degeneracy lines and tighten constraints on astrophysical parameters.}
    \label{fig:fisher_degeneracies}
\end{figure}

Fig.~\ref{fig:PS_Corner_plot} shows that joint posteriors for the three recovered parameters inferred from the spherically averaged power spectrum alone exhibit strong degeneracies. This is expected for a two-point statistic that averages over some physical information: correlations between parameters limit how well each can be constrained. Combining statistics with different degeneracy directions helps to break these correlations, since each probes complementary information. As illustrated in Fig.~\ref{fig:fisher_degeneracies}, the statistics considered here yield distinct parameter-correlation slopes. Using them in tandem -- for example, combining the bispectrum with the power spectrum \citep{WatkinsonGreig_2022,TiwariShaw_2022,Mahida_2025} -- can therefore tighten constraints relative to any single statistic, although inter-statistic correlations must be accounted for when they encode overlapping physical information (see Sec.~\ref{sec:combine_sum}).

The Fisher results presented in this section are not direct forecasts of the parameter constraints achievable with SKA-Low data. The analysis relies on idealised simulations: foregrounds and systematics are neglected, and instrumental noise enters only through the data covariance, even though it contributes to the observed signal. For WST, for instance, computing derivatives on observed rather than clean maps (i.e. after adding thermal noise and beam smoothing) weakens parameter constraints, motivating further work on the constraining power of these statistics under realistic observing conditions.
Finally, the potential of pixel-based and beyond-power-spectrum statistics extends beyond Fisher forecasts of a few astrophysical parameters on a single LC. Statistics such as the TCF and bubble-size distributions can probe the evolution and growth of ionised regions during the EoR, providing additional insight that we have not explored here.

\section{Towards maximising the information extraction}\label{sec:max_info}

While the summary statistics discussed in the previous sections offer significant constraining power, they are not guaranteed to be `sufficient' statistics that capture all the available information. This section delves into methods designed to capture this information more completely, either through the construction of optimal summaries (Sec. \ref{sec:info_max}), the combination of multiple complementary statistics (Sec. \ref{sec:combine_sum}), or by inferring parameters directly from the signal field (Sec. \ref{sec:fieldlevel}).

\subsection{Information maximising statistics}\label{sec:info_max}

An ideal approach to maximise the information extraction is to find 'sufficient' or 'optimal' summary statistics. The goal is to compress the high-dimensional raw data (like an image or visibility set) into a low-dimensional vector—ideally with the same number of components as the model parameters ($n \sim 10$)—while retaining all the physical information necessary for inference. This remains an open challenge, but machine learning offers several promising strategies.
One set of strategies involves training a network to explicitly optimise a statistical measure of information. For example, Information Maximizing Neural Networks (IMNNs) are designed to find a data compression that maximises the Fisher information matrix \citep{Charnock18,Prelogovic24}. While powerful, this approach requires defining a fiducial model for the Fisher matrix calculation. An alternative is to train a network to maximise the mutual information between the parameters and the learned summaries \citep{Jeffrey21,CerardiGiri_2025}.
Other approaches learn the summary statistics implicitly as part of a larger inference task. These include networks trained to recover parameter values directly, or summaries that are trained jointly with an inference network \citep{Neutsch:2022, Schosser:2024}. Task-independent methods, such as contrastive learning, can also find optimal data representations \citep{Ore:2025}. A crucial point is that even a perfectly sufficient statistic still has a stochastic likelihood (due to noise and sample variance) that must be modelled. Since this likelihood is often intractable, SBI (see Sec.~\ref{sec:inf_comp}) is a necessary tool to learn it, even when using these optimal summaries.

\subsection{Combining summary statistics}\label{sec:combine_sum}

Discovering sufficient statistics is challenging, so a simpler approach is combining summary statistics within Bayesian inference to increase information content. As we likely do not know the analytical form for the likelihood of these different types of summary statistics, and as they are likely not fully independent from each other, SBI should be used to obtain either the likelihood or directly the posterior. Several studies have shown applications of this principle, and obtained tighter constraints combining multiple summary statistics \citep{Prelogovic24,Semelin24,Schosser:2025,CerardiGiri_2025}.
For neural likelihood estimation, one partial limitation of this approach is the increased dimensionality of the combined statistics. 
This issue can be alleviated by using neural posterior estimator networks, whose outputs scale as the number of parameters that are targeted by inference, not as the dimensionality of the summary statistics.

\subsection{Field-level inference}
\label{sec:fieldlevel}

The most natural option to extract as much information as possible from a 21\,cm brightness temperature LC is to avoid reducing it to summary statistics. This idea forms the basis of field-level inference, where the model parameters include not only an astrophysical model of reionisation, but also the underlying matter distribution on a pixel-by-pixel basis. Such methods have initially been applied to galaxy surveys, in order to recover a distribution of the initial conditions of our Universe \citep[e.g., ][]{JascheKitaura_2010}. For example, \citet{Nguyen_2024} found that field-level inference applied to the dark matter halo field yielded a factor five tightening of constraints on the matter clustering amplitude $\sigma_{8}$.
Applying field-level inference to 21\,cm data during reionisation is difficult since increasingly large ionised bubbles obscure information as reionisation progresses. Despite this fact, field-level methods have already shown significant promise in 21\,cm cosmology. \citet{villanueva_2020} trained a U-Net to map 21\,cm images to density fields at $z \gtrsim 10$, and \citet{zhou2023reconstruction} achieved RMS errors $\lesssim7\%$ using maximum-a-posteriori (MAP) matter density field reconstructions that combine 21\,cm and CO intensity maps to compensate for the information missing inside ionised bubbles. \citet{zhao2023simulationbased} further applied SBI on 21\,cm LCs to constrain astrophysical and cosmological parameters. \citet{chen2025fieldlevelreconstructionforegroundcontaminated21cm} showed that field-level inference can accurately reconstruct 
modes with $k\lesssim0.8~h~\mathrm{Mpc}^{-1}$ in regions with ionization fractions $\lesssim60\%$, while simultaneously reducing the uncertainty on astrophysical parameters.
Recently, \citet{Kern_2025} presented a field-level inference framework that jointly estimates posteriors for the 21\,cm signal, foregrounds, and instrumental systematics, laying the groundwork for future SBI that simultaneously accounts for astrophysical signals and observational systematics. With continued improvements in high-dimensional samplers, such as the {\sc jax}-based \citep{jax2018github}, {\sc numpyro} \citep{phan2019composable} and {\sc Turing.jl} \citep{10.1145/3711897}, and increasingly capable AI architectures, field-level inference's ability to unlock the full information content of localized 21\,cm measurements is likely to grow substantially in the coming years.

\section{Discussion}\label{sec:discussion}

Inference methods and results in this chapter demonstrate clear progress in constraining cosmic-dawn and reionisation physics for the SKA era. Across modelling frameworks—from analytic approaches to complex numerical simulations—the shared aim is to capture the essential CD–EoR physics in a form that enables robust interpretation of SKA data. SKA-Low’s constraining power will depend not only on its sensitivity, but also on how well we model, compress, and analyse its observations. The shift from relying solely on the PS to incorporating higher-order and morphological statistics marks the transition from first detection to extracting maximal information. This, in turn, demands careful treatment of instrumental effects, foregrounds, and inference methodology. The following subsections examine these issues: the impact of modelling uncertainties (Sec.~\ref{sec:discussion_modelling}) and the broader prospects for cosmological and astrophysical inference with SKA-Low.

\subsection{Impact of modelling uncertainty on the constraints}  \label{sec:discussion_modelling}
\begin{figure}[h!]
    \centering
    \includegraphics[width=0.70\textwidth]{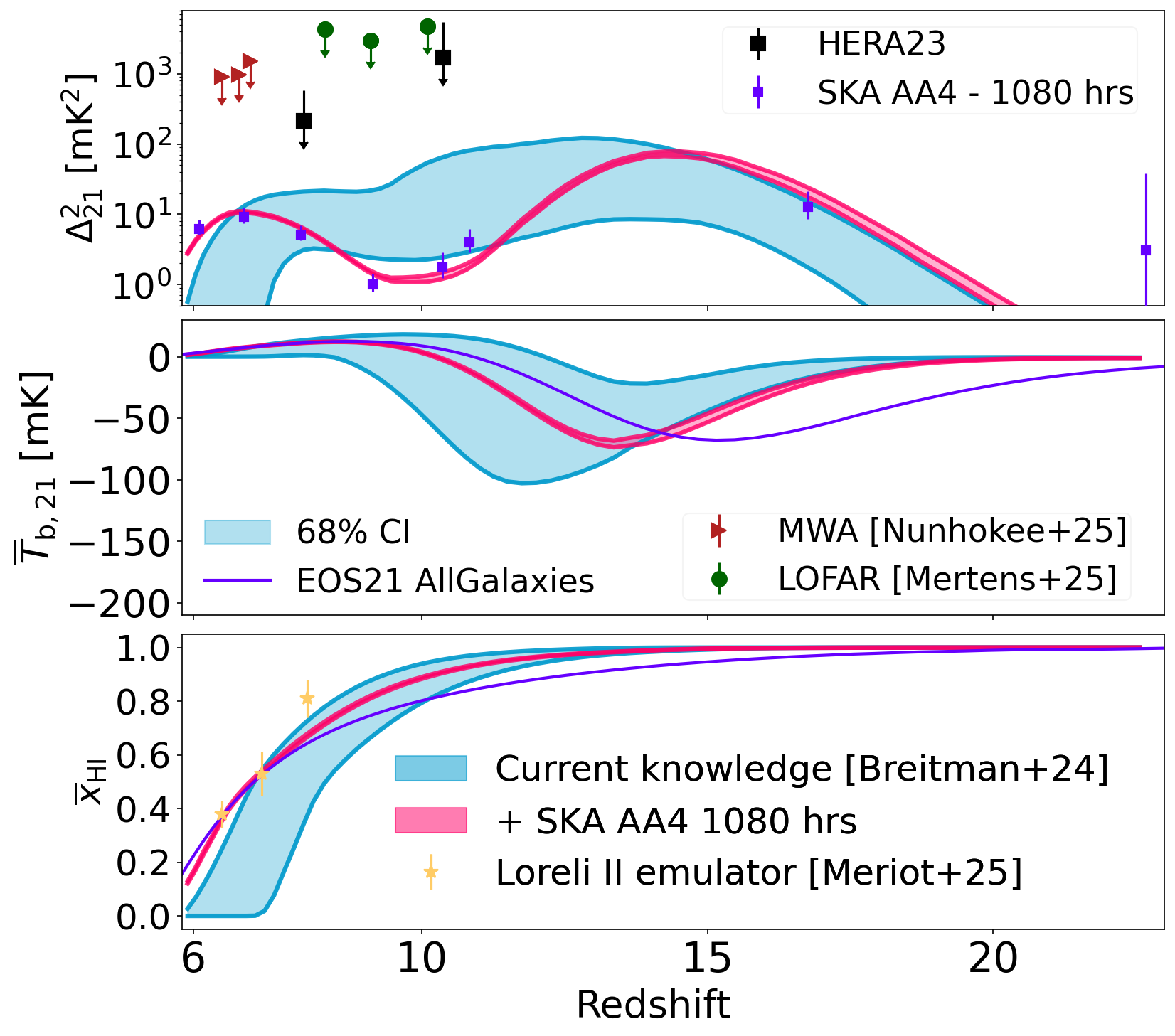}
    \caption{Same as Fig. \ref{fig:inference_results}, but including mock SKA observations from EOS21 \citep{Munoz22}, shown as purple squares (top panel) and purple lines (other panels). The magenta posterior replaces the HERA upper limits with these mock SKA detections, while the golden stars show results obtained with the Loreli II emulator \citep{Meriot24b}. Posterior recovery of the EoR history is poor because both emulators were trained on a different galaxy model than that of the mock SKA observation.
    }
\label{fig:inference_results_modelling}
\end{figure}

Extracting robust astrophysical and cosmological constraints from upcoming SKA 21\,cm observations requires accurate signal models. Because CD–EoR physics spans a vast dynamic range from ISM to IGM scales, no single approach can capture all relevant processes with both accuracy and efficiency (Sec.~\ref{sec:modelling}). Full hydrodynamic RT simulations provide the most physically consistent modelling but are computationally expensive and restricted to relatively small volumes (Sec.~\ref{sec:num}), whereas semi-numerical methods efficiently explore larger parameter spaces, and reproduce large-scale RT statistics within $\sim 10\%$ accuracy (Sec.~\ref{sec:sem_num}). Each approach nonetheless introduces biases and model misspecification that should be quantified or marginalised over during inference.

These modelling uncertainties propagate directly into inference. Bayesian analyses of the 21\,cm power spectrum typically require $\sim 10^6$ forward-model evaluations, too costly for direct simulation and motivating fast emulators (Sec.~\ref{sec:emu}). Tests with different emulator families show that mismatches between the training models and the `true' Universe can bias recovered ionisation histories even when the power-spectrum fit appears reasonable. Statistical choices compound this: power spectra and Gaussian likelihoods remain the primary tools to characterise the EoR (Sec.~\ref{sec:PS}) but do not capture the full complexity of the data, while foregrounds, systematics, and LC evolution imprint additional non-Gaussian structure. Statistics beyond the power spectrum (Sec.~\ref{sec:higher_order}--\ref{sec:Fisher_Discussion}) and flexible inference methods such as SBI offer alternatives that more robustly handle such effects (Sec.~\ref{sec:inf_comp}). Forecasts that neglect modelling uncertainties or assume perfect foreground removal, simplified noise, or idealised statistics therefore overestimate constraining power and bias parameter inferences. See \citet{GreigMesinger_2015} and \citet{SchneiderSchaeffer_2023} for discussions of their impact on astrophysical and cosmological inference, respectively. In practice, constraints degrade when using realistically instrumentally degraded maps, underscoring the need to validate models, emulators, and inference pipelines against the full range of physical and observational effects.

We illustrate the effects of model misspecification in Fig.~\ref{fig:inference_results_modelling} (cf. Fig.~\ref{fig:inference_results}) using a mock SKA AA4 observation based on the EOS21 model \citep{Munoz22} (purple) and two emulators trained on different physics and/or simulators. (i) {\sc 21\,cmEMUv1} \citep{BreitmanMesinger_2024} (pink) is trained on simulations from the {\sc 21\,cmFASTv3} semi-numerical simulator \citep{Murray2020}, which also produced EOS21, but its training set includes only Pop~II stars, whereas EOS21 additionally models Pop~III stars. (ii) {\sc Loreli II} \citep{Meriot24b} (yellow) is trained on the {\sc Licorice} database of radiative-transfer simulations. The pink posterior on the power spectrum matches the observations reasonably well, but the true $T_{b,21}$ and $x_{\mathrm{HI}}$ histories lie outside the corresponding posteriors. Thus, even though EOS21 and {\sc 21\,cmEMUv1} share the same simulator, differences in the physical model are sufficient to bias the inference when model uncertainty is not accounted for in the setup. Repeating the inference with SBI using an emulator trained on the {\sc Loreli II} database of {\sc Licorice} simulations \citep{Meriot24b, Meriot24}, in which both the physics and the simulator differ from the mock observation, further emphasises this: the yellow points yield a posterior on the neutral fraction that diverges even more strongly from the true history at moderate redshifts. Comparing this neutral-fraction recovery with Fig.~\ref{fig:inference_results}, where the mock and emulator share the same model and the EoR history is recovered to percent-level precision, highlights the importance of controlling model misspecification in SKA 21\,cm inference.

\subsection{Cosmological inference} \label{sec:discussion_cosmology}

The 21\,cm signal depends directly on the cosmological content and large-scale matter distribution of the Universe, making it a sensitive probe of cosmology. Multiple studies show that it can be used to constrain cosmological parameters, inflationary models, the nature of dark matter, and cosmologies beyond $\Lambda$CDM. Within the $\Lambda$CDM paradigm, 21\,cm power-spectrum measurements from precursors such as MWA and HERA and pathfinders such as LOFAR can already constrain cosmological parameters \citep{McQuinn2006,Mao2008,Pritchard2008,Clesse2012,Liu2016,Liu2016b,kern2017emulating,Munoz2019,Munoz2019b,Cain2020,Sarkar2023,Flitter2024,Acharya_2025,maity_2025}, and SKA forecasts using the power spectrum have been made for the six standard cosmological parameters \citep{Mao2008,SchneiderSchaeffer_2023}. Tomographic 21\,cm imaging with SKA will further improve constraints \citep{Mao2008,Hassan2020,Neutsch:2022,Thelie2025}; for instance, \citet{Hassan2020} and \citet{Neutsch:2022} use deep-learning techniques on 21\,cm images to recover cosmological parameters. Several studies propose standard rulers during CD-EoR \citep{Munoz2019,Munoz2019b,Cain2020,Sarkar2023,Thelie2025}: velocity-induced acoustic oscillations (VAOs) provide a calibrated standard ruler in the 21\,cm PS that constrains the expansion history through the Hubble rate $H$ at cosmic-dawn redshifts \citep{Munoz2019b,Sarkar2023}, while ionised bubbles act as uncalibrated standard spheres in an Alcock–Paczy\'nski test, forecasting constraints on the product $D_{\rm A}H$ (with $D_{\rm A}$ the comoving angular-diameter distance) during reionisation with future SKA 21\,cm images \citep{Thelie2025}. 

21 cm observations can also constrain inflationary physics. For example, while the 21 cm PS can probe the inflationary potential and slow-roll parameters \citep{Barger2009,Adshead2011}, and primordial gravitational waves \citep{he2024inverse}, the trispectrum can test primordial non-Gaussianity \citep{Cooray2008}. Furthermore, the 21 cm signal holds information about the nature of dark matter. The redshift evolution of the signal is particularly powerful for testing exotic dark matter models, as deviations from cold dark matter leave distinctive imprints on large-scale structure formation and the cosmic thermal history \citep[e.g.,][]{Schneider_2018,GiriSchneider_2022}. Various dark matter candidates have been studied, including sterile neutrinos, decaying dark matter and annihilating dark matter \citep[e.g.,][]{Furlanetto2006,Valdes2007,LopezHonorez2016}, warm dark matter \citep[e.g.,][]{Sitwell2014,saxena2020impact}, millicharged dark matter \citep[e.g.,][]{Munoz2015,Munoz2018}, fuzzy dark matter \citep[e.g.,][]{Jones2021,NebrinGhara_2019}, mixed fermionic–bosonic dark matter \citep[e.g.,][]{GiriSchneider_2022}, and scattering dark matter \citep{Flitter2024}.
Additionally. modified-gravity signatures have been predicted in both 21\,cm images and the PS \citep{Hall2013,Brax2013}, and forecasts exist for general modified-gravity models \citep{Heneka:2018ins}. Coupled-quintessence and early–dark-energy scenarios can also be constrained with the 21\,cm global signal and PS \citep{Liu2020,AdiFlitter_2025}. 
Extracting this cosmological information is complicated by degeneracies between cosmology and astrophysics, and unless such degeneracies are overcome, cosmological constraints will remain limited \citep[e.g.,][]{LopezHonorez2016,Agius2025}. Synergies with complementary observables, such as high-redshift UV luminosity functions and quasar spectra, will be helpful in this study \citep[e.g.][]{Dhandha2025,Acharya_2025}.

\section{Conclusion}\label{sec:conclusion}

The SKA will mark a significant milestone in our ability to infer the physics of the early Universe from the 21\,cm signal. With its unparalleled sensitivity and resolution, SKA-Low will enable us to progress from upper limits and tentative detections to genuine quantitative understanding of the first billion years of cosmic history. Combining observed data with the novel inference techniques reviewed in this chapter, we will link the 21\,cm signal directly to the properties of the first galaxies and the evolution of the IGM.
The prospects for inference are substantial. We will be able to explore the astrophysics of early sources and the underlying cosmology within a single framework, using a range of complementary summary statistics and field-level analyses. These advances will constrain models of reionisation and heating, test fundamental cosmological scenarios, and potentially reveal new physics in the high-redshift Universe.
SKA-Low represents both the culmination of years of theoretical and methodological development and the beginning of a new observational era. Its observations will challenge our models and assumptions about the early Universe. Inference from SKA-Low data will be a process of discovery, transforming the faint cosmological signal into a detailed understanding of how the first stars and galaxies emerged and reionised the IGM.

\section{Author contributions}\label{sec:author_contribution}

S.K.Giri, A.Gorce, I.Hothi, A.K.Shaw, and C.M.Trott coordinated this chapter, including organising contributions, structuring the narrative, and substantively editing the text throughout. I.Hothi developed the database used to run the Fisher forecasts. S.K.Giri, A.Gorce, and I.Hothi wrote a template analysis pipeline to be adapted for the different statistics tested in the Fisher forecasts (Secs.~\ref{sec:PS_fisher}, \ref{sec:Fisher_Discussion}, and Appendix~\ref{sec:fisher_framework}). The analysis pipeline and individual contributions are publicly available at \url{https://github.com/sambit-giri/ska_eor_inference}. S.G.Murray provided a final editorial pass across the chapter, improving readability with review, stylistic suggestions, and edits. C.Heneka provided comments across the chapter. We give the contributions of the authors to every section below.
\paragraph{Sec.~\ref{sec:intro}} C.M.Trott and I.Hothi wrote this section.
\paragraph{Sec.~\ref{sec:modelling}.} A.Schneider wrote Sec.~\ref{sec:analytical} (analytical modelling). Sec.~\ref{sec:sem_num} on semi-numerical simulations was primarily written by S.Majumdar, A.Acharya, and B.Maity. Sec.~\ref{sec:num} was first drafted by M.Bianco, who also produced Fig.~\ref{fig:lc_compare}, with contributions from S.K.Giri. D.Breitman primarily wrote Sec.~\ref{sec:emu}, and A.Tripathi, Y.Mahida, A.Datta, and S.Majumdar wrote the third paragraph of this subsection.
\paragraph{Sec.~\ref{sec:PS_inf}} N.Kern, S.K.Giri, A.Gorce, and I.Hothi wrote Sec.~\ref{sec:PS} and produced Figs.~\ref{fig:PS_fiducial} and~\ref{fig:PS_Corner_plot}, with A.K.Shaw and R.Mondal writing the MAPS subsection (Sec.~\ref{subsec:maps}). D.Breitman and R.M\'eriot contributed to Sec.~\ref{sec:inf_comp}, with H.Shimabukuro writing the Direct Parameter Prediction paragraph, additional contributions from A.Tripathi, and S.K.Giri writing the last paragraph. D.Breitman produced Fig.~\ref{fig:inference_results}.
\paragraph{Sec.~\ref{sec:beyond_PS}} The bispectrum section (Sec.~\ref{sec:bispectrum}) was primarily written by L.Noble, A.K.Shaw, Y.Mahida, S.Majumdar, and A.Gorce. S.K.Giri wrote Sec.~\ref{subsec:posdep_ps} (position-dependent power spectrum) and I.Hothi wrote Sec.~\ref{sec:scattering} (scattering transform). A.Gorce and C.M.Trott wrote Sec.~\ref{subsec:moments} (moments). In Sec.~\ref{sec:bubble_stats}, the description of bubble sizes was mainly written by I.Georgiev and S.K.Giri, with contributions from S.K.Pal, S.Dasgupta, S.Bag, A.Datta, and S.Majumdar. A.Kapahtia and S.K.Giri wrote the initial text for the topological measures. Sec.~\ref{sec:time_fields} on reionisation time was mainly written by J.Hiegel with contributions from E.Th\'elie. For Sec.~\ref{sec:Fisher_Discussion}: L.Noble, Y.Mahida, A.K.Shaw, and S.Majumdar performed the bispectrum Fisher analysis; A.K.Shaw and R.Mondal performed the MAPS Fisher analysis; L.Crascall-Kennedy ran the Fisher forecasts for the Triangle Correlation Function with contributions from M.Douspis and A.Gorce; and S.K.Giri ran the forecast for the position-dependent power spectrum. A.Gorce, I.Hothi, and S.K.Giri wrote the summary and discussion of the forecast outputs. A.Gorce produced Fig.~\ref{fig:fisher_degeneracies}.
\paragraph{Sec.~\ref{sec:max_info}} B.Semelin wrote Secs.~\ref{sec:info_max} and \ref{sec:combine_sum}. S.Berger wrote Sec.~\ref{sec:fieldlevel}. 
\paragraph{Sec.~\ref{sec:discussion}} C.Heneka wrote Sec.~\ref{sec:discussion_modelling} with contributions from R.M\'eriot and D.Breitman. R.M\'eriot ran Bayesian inferences for Fig.~\ref{fig:inference_results_modelling} in collaboration with D.Breitman. E.Th\'elie wrote Sec.~\ref{sec:discussion_cosmology}.
\paragraph{Sec.~\ref{sec:conclusion}} I.Hothi wrote this section with contributions from A.Gorce and S.K.Giri.

We also thank the anonymous reviewer for their constructive feedback.

\bibliographystyle{abbrvnat-maxbibnames4}
\bibliography{chapter} 

\appendix

\section{Fisher inference framework}\label{sec:fisher_framework}

\subsection{Fisher forecast}

Given a posterior distribution $P(\theta\vert\mathrm{data})$ for model parameters $\theta$, the Fisher-matrix components are given by \citep[e.g.,][]{kay93, Repp_2015}:
\begin{equation}
\mathcal{F}_{i j}=-\left\langle\frac{\partial^2 \ln P}{\partial \theta_i \partial \theta_j}\right\rangle.
\end{equation}
The Cramér-Rao theorem states that any unbiased estimator for the parameters will produce a covariance matrix that is no more accurate than $\mathcal{F}^{-1}$ \citep{cramer, rao_1945}: thus the Fisher matrix can be used to estimate the minimum uncertainties of parameters given observations.

Here, the observable is the chosen statistic (e.g., the spherical power spectrum) computed over a range of bins, for a given LC. Let us gather these data points in an observable vector $\mathbf{o}$ of dimension $n$. We have $\mathbf{\Pi}$ the data covariance matrix of dimension ($n$, $n$). Given the parameter set $\theta$ -- here, 21\,cmFAST model parameters (see next section), and assuming the posterior distribution can be described as Gaussian, we define the $(i, j)$ element of the Fisher matrix as  \citep[e.g.,][]{kay93, Repp_2015, Abinash_infer2020}
\begin{equation}
\mathcal{F}_{ij}= \frac{\partial \mathbf{o}^T}{\partial \theta_i}\mathbf{\Pi}^{-1}\frac{\partial \mathbf{o}}{\partial \theta_j}.
\end{equation}

In these ideal conditions, the inverse of the Fisher matrix is the covariance matrix $\mathcal{C}=\mathcal{F}^{-1}$ and the forecast uncertainty on the $i$ th parameter is $\sigma(\theta_i)=\sqrt{\mathcal{C}_{ii}}$.

\subsection{Description of the simulations}\label{sec:fisher_simulations}

The Fisher matrices were calculated using {\fontfamily{cmtt}\selectfont 21\,cmFAST} simulations, which were originally simulated for this purpose in \citet{HothiAllys_2024}. {\fontfamily{cmtt}\selectfont
21\,cmFAST}\footnote{https://21\,cmfast.readthedocs.io/en/latest/} \citep{MesingerFurlanetto_2011,Murray2020} is a widely used semi-numerical simulation code of the 21\,cm signal during the EoR (see Sec.~\ref{sec:sem_num}). 
To compute the Fisher matrices, one calculates the derivative, with respect to a given parameter, by modifying the parameter about its fiducial value. We use the simulation developed in \citet{HothiAllys_2024} where they varied three astrophysical parameters. The first parameter varied is $T_{Vir}$, which is minimum Virial temperature needed for halos to host star-forming galaxies. The second parameter is $R_{\rm{max}}$, which signifies the maximum distance a photon can travel within the simulated field before encountering a neutral hydrogen atom; this is loosely related to the mean free path. The final parameter is the ionising efficiency of galaxies $\zeta$, which depends on the properties of the these galaxies such as ionising photon escape and star formation rate \citep[see eq.~2 in][]{GreigMesinger_2015}.

For the Fisher analysis, we varied each parameter about its fiducial value as: 1) $T_{\rm Vir}=50000\pm 5000$ K, 2) $R_{\rm Max}=15\pm 5$ Mpc, and 3) $\zeta=30\pm 5$.
Each of the simulated LCs has a transverse extent of 200 $h^{-1} \rm{Mpc}$ and consists of 256 pixels per side in each frequency channel. The LCs span a redshift range from $z = 8.82$ (144.60 MHz) to $z = 9.33$ (137.46 MHz), comprising 128 frequency channels.
For each parameter change, 
400 simulations were run by changing the initial conditions. The covariance matrix is estimated from an ensemble of 400 simulations run, that were run through \textsc{tools21\,cm} (see Sec. \ref{sec:fisher_obs_setup}), using only the fiducial parameter values. This method ensures that our covariance estimate properly accounts for sample variance, as well as observational effects such as noise, which are discussed next.

\subsection{Observational setup} \label{sec:fisher_obs_setup}

The spatial distribution of the 21\,cm signal is measured by a radio interferometer, which consists of multiple antenna stations. Each pair of antennas forms a baseline, and different baselines probe different spatial scales by measuring the corresponding visibilities. We simulate SKA-Low observations following the method described in \citet{GiriMellema_2018a}, implemented in the \textsc{tools21cm}\footnote{\url{https://tools21cm.readthedocs.io/}} package. This package can generate noise maps, enabling the study of various summary statistics derived from 21\,cm data.

The root-mean-square (RMS) of the instrumental noise is given by \citep{GiriMellema_2018a}:
\begin{equation}
\mathrm{RMS} = \frac{2k_\mathrm{B}^2 T_\mathrm{sys}^2}{\epsilon A_\mathrm{D}^2 t_\mathrm{int} \Delta \nu} \ ,
\end{equation}
where $k_\mathrm{B}$ is the Boltzmann constant, $A_\mathrm{D}$ is the collecting area of each station, $\Delta\nu$ is the frequency resolution, and $t_\mathrm{int}$ is the integration time. The efficiency factor $\epsilon$ is unity for frequencies $\nu > 110$ MHz. The system temperature is modelled as $T_\mathrm{sys} (\nu)= \left[100+60\left(\frac{\nu}{300\mathrm{MHz}}\right)^{-2.55}\right]$ K. The sensitivity of a radio interferometer depends on the number density of baselines at a given scale. SKA-Low, currently under construction, will be deployed in multiple phases, including AA* and AA4, each with different antenna distributions. The planned antenna locations for these phases are provided by SKAO through the public \textsc{ska-ost-array-config}\footnote{\url{https://gitlab.com/ska-telescope/ost/ska-ost-array-config}} package, which is interfaced by \textsc{tools21\,cm}. Our simulations account for Earth’s rotation to fully sample the $uv$-plane.

We assume an observing strategy of 6 hours per day with $t_\mathrm{int} = 10$s. Longer total integration times (e.g., 1000 h) are obtained by combining data from multiple days. The baseline density drops significantly beyond 2 km; to avoid spurious features from incomplete $uv$ coverage, we smooth the signal at scales corresponding to baselines longer than this. The frequency resolution $\Delta\nu$ is chosen to remove comparable small-scale modes along the line of sight.

\end{document}